# What Lives? A meta-analysis of diverse opinions on the definition of life

Reed Bender[1#], Karina Kofman[2#], Blaise Agüera y Arcas[3], and Michael Levin[4,5*]

Affiliations:

[1] Attune Intelligence, LLC
[2] Faculty of Dentistry, University of Toronto, Toronto, Canada
[3] Paradigms of Intelligence Team, Google
[4] Allen Discovery Center at Tufts University
[5] Wyss Institute for Biologically Inspired Engineering at Harvard University

[#] Co-first authors

* Author for Correspondence
200 Boston Ave.
Suite 4600
Medford, MA 02155
email: michael.levin@tufts.edu

## Abstract

The question of "what is life?" has challenged scientists and philosophers for centuries, producing an array of definitions that reflect both the mystery of its emergence and the diversity of disciplinary perspectives brought to bear on the question. Despite significant progress in our understanding of biological systems, psychology, computation, and information theory, no single definition for life has yet achieved universal acceptance. This challenge becomes increasingly urgent as advances in synthetic biology, artificial intelligence, and astrobiology challenge our traditional conceptions of what it means to be alive. We undertook a methodological approach that leverages large language models (LLMs) to analyze a set of definitions of life provided by a curated set of cross-disciplinary experts. We used a novel pairwise correlation analysis to map the definitions into distinct feature vectors, followed by agglomerative clustering, intra-cluster semantic analysis, and t-SNE projection to reveal underlying conceptual archetypes. This methodology revealed a continuous landscape of the themes relating to the definition of life, suggesting that what has historically been approached as a binary taxonomic problem should be instead conceived as differentiated perspectives within a unified conceptual latent space. We offer a new methodological bridge between reductionist and holistic approaches to fundamental questions in science and philosophy, demonstrating how computational semantic analysis can reveal conceptual patterns across disciplinary boundaries, and opening similar pathways for addressing other contested definitional territories across the sciences.

## 1. Introduction

The challenge of defining life extends beyond semantics; it reveals fundamental epistemological differences in how experts across disciplines conceptualize and investigate the boundaries between living and nonliving systems. Biologists often emphasize metabolic processes, reproduction, and cellular organization [1-6], while physicists may focus on thermodynamic properties, energy utilization, and entropy reduction [7-9]. Computer scientists and complexity theorists might alternatively prioritize information processing, self-organization, and emergent complexity [10-14].

These divergent perspectives, shaped by distinct methodological approaches and theoretical frameworks, have historically been difficult to reconcile. This definitional challenge becomes increasingly urgent as advances in synthetic biology [15], artificial intelligence [16], and astrobiology [17] continue to challenge what it means to be alive, which has not only scientific but ethical and regulatory implications.

Some have advocated for definitional pluralism — the acceptance of multiple, contextually appropriate definitions rather than pursuing a single universal concept [18,19]. This pragmatic approach acknowledges the value of specialized definitions within differentiated research contexts. However, it leaves unanswered the question of whether deeper patterns of coherence might exist beneath these apparent disciplinary divisions, and whether a more integrated understanding is possible without sacrificing disciplinary perspectives.

Others have rejected binary categorizations entirely, criticizing attempts to draw arbitrary lines between the living and nonliving. These perspectives reframe the conversation around characterizing life as existing along multiple continuous dimensions rather than as a categorical distinction [20-23]. More radically, some scholars question the utility of defining life at all, suggesting that such attempts are either conceptually futile or entirely misguided [24-26].

We sought to survey a set of contemporary scholars for their definition of life and then use both manual and computational semantic analysis to examine the resulting dataset. Our research addressed the definitional ambiguity through a methodological approach that leverages large language models (LLMs) to systematically analyze 68 expert-provided answers to the question of "what is life?" We implemented an LLM-derived pairwise correlation analysis to quantitatively map the collective set of responses into a collective correlation matrix, followed by agglomerative clustering and dimensionality reduction to project those encoded and clustered respondent feature vectors onto a 2-dimensional plane. To further interpret this visualized semantic space, we leveraged LLMs to compare intra-cluster definitions of life and to generate a consensus definition of life that is representative of each cluster.

By systematically mapping how expert perspectives vary across perceptual frames, we aim to provide a multidimensional framework for understanding "aliveness" that transcends traditional categorical distinctions to guide ethical and conceptual decisions in our rapidly evolving technological landscape.

## 2. Background

**Historical Conceptions of Life**

Long before the emergence of modern scientific frameworks, ancient civilizations sought to characterize the animating force which gives rise to life. Rather than providing naturalistic explanations, these ancient cultures relied on supernatural abstractions such as Prana, Qi, Ka, Pneuma, and Ruach in the Vedic, Chinese, Egyptian, Greek, and Hebrew traditions respectively. These concepts, while diverse in their cultural origins, shared a common attempt to articulate an invisible principle that separated animate from inanimate matter within metaphorical abstractions.

Greek philosophy transformed these symbolic conceptions into the first systematic definitional frameworks of the soul and its relation to life. Plato's (428 – 348 BCE) cosmology in the Timaeus characterizes the cosmos itself as a living and intelligent being, with its motions governed by cognition rather than mechanical causation, thereby unifying the soul as both the principle of cognition and the principle of life for the first time [27]. For Plato, the differentiated life of the microcosm was a mirror of the expression of life at the scale of the macrocosm.

> *"The permanent value of the Timaeus rests in its successful presentation of a cosmological basis for a theoretical and practical ethics for mankind. Its theme is life, the generating principle of life — not merely the life of one man or even of humanity, but the genesis τοῦ παντός[1], a fit subject for any philosopher [28]."*

Aristotle (384 – 322 BCE) then advanced this framework through his teleological approach in De Anima, further characterizing life by its manifested functional properties. His biological framework presents a nested hierarchy of capacities, where the vegetative soul (nutrition and reproduction) is present in all living things, the sensitive soul (perception and movement) in animals, and the rational soul uniquely in humans [29]. His emphasis on intrinsic purpose (telos) established the first ontologically robust definition of life as a self-actualizing system, where the most natural act of living things is "the production of another like itself, an animal producing an animal, a plant a plant, in order that, as far as nature allows, it may partake in the eternal and divine" [30]. For Aristotle, the goal towards which all living things strive is to "do whatsoever their nature renders possible," establishing living things as fundamentally self-actualizing systems with intrinsic teleology.

The transition from Greek thought to our modern materialistic conceptions of life was mediated by medieval and Renaissance thinkers who preserved teleological frameworks while introducing increasingly material explanations. Galen (129 – 216 CE) extended Aristotelian biological teleology through empirical anatomy, demonstrating the purposive construction of every anatomical feature [31]. Renaissance alchemists, particularly Paracelsus (1493 – 1541), subsequently developed hybrid frameworks that viewed organisms as chemical laboratories governed by an "archeus" or inner alchemist — a concept that maintained a sense of teleological functionalism while increasingly describing life by its biochemical form [32]. These "chemical philosophies" represented an important conceptual evolution [33], maintaining the notion that living systems operated toward intrinsic ends while simultaneously beginning to articulate material processes underlying vital functions. This historical progression established a foundational tension between teleological and mechanistic perspectives that continues to challenge contemporary efforts to define life.

[1] Greek *Tou Pantos*, "of the all" or "of the universe/cosmos," the gestalt of all creation.

During the Enlightenment, René Descartes (1596 – 1650) conceptualized the human as an assemblage of parts that together operated as a machine, dualistically distinct from the mind which governed its function. This mechanistic approach to the expression of life fundamentally reoriented its conception away from its teleological roots, redirecting focus towards its mechanistic processes.

> "*Thus, I say, when you reflect on how these functions follow completely naturally in this machine solely from the disposition of the organs, no more nor less than those of a clock or other automaton from its counterweights and wheels, then it is not necessary to conceive on this account any other vegetative soul, nor sensitive one, nor any other principle of motion and life, than its blood and animal spirits, agitated by the heat of the continually burning fire in the heart, and which is of the same nature as those fires found in inanimate bodies* [34]."

While Descartes described life as a duality of immaterial mind and its mechanistic body, his contemporaries developed thoroughly materialistic philosophies from the seed of his work. Thomas Hobbes (1588 – 1679) advanced perhaps the most comprehensive materialistic framework of this period, rejecting Cartesian dualism entirely and arguing that all phenomena characteristic of life could be reduced to matter in motion. In conceptualizing life, Hobbes rejected prior teleologic or metaphysical conceptions in favor of the view that biological functions are purely mechanical processes.

> "*For seeing life is but a motion of Limbs, the beginning whereof is in some principal part within; why may we not say, that all Automata (Engines that move themselves by springs and wheels as doth a watch) have an artificial life? For what is the Heart, but a Spring; and the Nerves, but so many Strings; and the Joints, but so many Wheeles, giving motion to the whole Body, such as was intended by the Artificer?* [35]."

Throughout the Enlightenment, this mechanistic naturalization of life continued to evolve through philosophers such as Gassendi [36], Spinoza [37], and La Mettrie [38], culminating in a thoroughly materialist framework that would define our modern scientific age. Yet even as the philosophical landscape shifted decisively toward mechanistic explanations, the fundamental question of how to define life persisted. By the dawn of the 19th century, scientific inquiry had not resolved but rather reformulated this ancient problem: if life emerges from purely physical processes, what distinguishes living from non-living matter?

## Modern Conceptions of Life

The transition into modern scientific conceptions of life was catalyzed by a series of pivotal 19th century developments: cell theory identified the fundamental organizational unit of life [39]; Charles Darwin's theory of evolution by natural selection provided the first mechanistic framework for explaining life's complexity and adaptation [40]; the emergence of thermodynamics laid the physical foundation for understanding how living systems maintain order in spite of the Second Law of Thermodynamics [41].

These advances set the stage for Erwin Schrödinger's seminal 1944 lectures, "What is Life?", which framed organisms as thermodynamic systems that create and maintain order by generating entropy in their environment [7]. This thermodynamic perspective, while influential, immediately faced challenges: could a purely energetic characterization capture the qualitative difference between living and non-living?

The molecular revolution of the mid-20th century promised resolution. James Watson and Francis Crick's elucidation of DNA's double helix structure unveiled the physical basis of heredity [42], while subsequent discoveries in molecular biology revealed the central dogma of genetic information flow [43]. This mechanistic understanding of heredity spawned information-theoretic definitions, most notably by John von Neumann, who characterized life as a computational system capable of self-replication with inheritable variation [44].

Contemporary definitional approaches have proliferated across disciplines, diversifying the available perspectives from which one might define life. Thermodynamic frameworks emphasize energy flux and entropy production [45], while autopoietic theories, such as those developed by Maturana and Varela, focus on self-organization and boundary maintenance [46]. The Chemoton model by Gánti proposes minimal requirements including metabolism, information storage, and boundary control [47]. NASA's operational definition, adopted from a suggestion by Carl Sagan, represents a pragmatic definition of life for astrobiological research: "a self-sustaining chemical system capable of Darwinian evolution" [17].

Despite the complexification of life's definitional landscape, tension between perspectives persists. Thermodynamic approaches may capture energy dynamics while underspecifying organizational complexity. Information-theoretic models illuminate genetic coding but struggle with the semantics of biological meaning. Operational definitions encounter borderline cases: viruses, prions, and computational life forms that exhibit some but not all traditional life characteristics.

The emergence of artificial life, synthetic biology, and complex computational systems has further complicated definitional boundaries. Craig Venter's synthetic Mycoplasma cell demonstrated the feasibility of constructing a viable synthetic cell with a compressed genomic code [48]; Michael Levin has shown that embryonic frog cells can self-assemble into novel forms of biology when removed from their traditional physiological context [15]; artificial neural networks [49] and evolutionary algorithms [50] exhibit adaptation and reproduction without biological substrates. These developments challenge the assumption that life requires carbon-based chemistry, cellular organization, or evolutionary adaptation to exhibit characteristics of life.

Systems biology has attempted synthesis through hierarchical frameworks that integrate multiple organizational levels, from molecular networks to ecosystems [51].

However, even this integrative approach must contend with the fundamental question: which properties represent necessary versus sufficient conditions for life?

The proliferation of definitional frameworks reflects both scientific progress and persistent conceptual challenges in our understanding of life. As our empirical science deepens, the boundaries of life's definition paradoxically become less distinct. We conceptualize these ontological tensions as a semantic topology that individual disciplinary approaches have mapped only partially, awaiting the tools of computational systematic analysis to reveal the underlying continuities present within this latent space.

To begin to explore this space and develop methods for parsing the conceptual landscape of scientists' opinions in difficult, interdisciplinary domains, we undertook an AI-guided analysis of a set of modern scientists' definition of “Life”. Hand-picked thinkers in a range of disciplines were asked to define “Life” (or argue against the possibility of doing so) in three sentences or less. Our goal was to map out the structure and main drivers of the highly diverse set of responses and evaluate the utility of LLMs for helping to do so.

## A Priori Computational Text Clustering Methods

Our approach builds upon recent advances in AI-augmented clustering, consensus formation, and pairwise constraint modeling, applying these techniques to analyze conceptual patterns in the semantic space of our respondents' definitions of life.

LLMs have been previously demonstrated to achieve comparable or superior performance to state-of-the-art text embedding and clustering methods by asking the model to generate potential labels for a given dataset and then subsequently categorizing each sample with an appropriate label [52]. More commonly, researchers utilize LLMs to generate text embeddings that encode contextual meaning into dense floating-point vectors, where abstract linguistic patterns can be represented geometrically. Comparative analysis of LLM embeddings for clustering demonstrates how these models consistently capture semantic relationships encoded as geometric proximity in high-dimensional latent spaces [53].

Recent innovations have further enhanced these capabilities through interpretable k-means clustering and instruction-tuned feedback mechanisms. The k-LLMmeans algorithm utilizes LLMs to generate textual summaries as cluster centroids, capturing semantic nuances often lost when relying on the purely mathematical properties of k-means clustering [54]. The ClusterLLM framework introduces an LLM to improve embedded-text clustering by constructing triplet questions <does A better correspond to B than C>, where A, B and C are similar data points that belong to different clusters according to the original embedder [55]. Following this same paradigm, Liusie, et al., have demonstrated that LLMs performed significantly better at comparative assessment tasks (e.g. “Which summary is more coherent, A or B?”) than at absolute assessment tasks (e.g. “Provide a score between 1 and 10 that measures this summary's coherence.”) [56]. These advances demonstrate how LLMs can leverage both embedded semantic relationships and direct linguistic instruction to achieve superior clustering performance, especially when prompted in the form of pairwise comparative analysis. This capability for interpretable analysis through pairwise comparison makes LLMs particularly suitable for mapping the complex conceptual landscape of life's definitions.

These LLM capabilities – semantic clustering, pairwise comparative analysis, and interpretable pattern recognition – directly address the methodological challenges inherent in analyzing diverse definitions of life. Traditional approaches have struggled to systematically compare and integrate definitions across disciplinary boundaries. Our approach treats expert definitions as data points within a quantifiable semantic space, enabling computational analysis of patterns that resist conventional categorical frameworks. By applying these techniques to the 68 expert responses, we can map the conceptual relationships between different definitional approaches while maintaining the nuance and complexity of each perspective. This methodology transforms the question "what is life?" from a philosophical debate into a structured exploration of semantic patterns within expert discourse.

## 3. Methods

We developed a software architecture to analyze and cluster diverse definitions of life using large language models (LLMs) and unsupervised learning techniques. Our approach encompasses five key methodological components:

1. Expert curation and definition collection
2. Quantitative pairwise correlation analysis between definitions
3. Agglomerative clustering of the resulting correlation matrix
4. Thematic analysis of intra- and inter-cluster semantic patterns
5. t-SNE dimensionality reduction of correlation feature vectors to 2-dimensional space

Figure 1 shows the key computational steps taken to go from the raw pairwise correlation matrix, to a sorted and clustered matrix, to ultimately a 2-dimensional t-SNE projection of the definitional space.

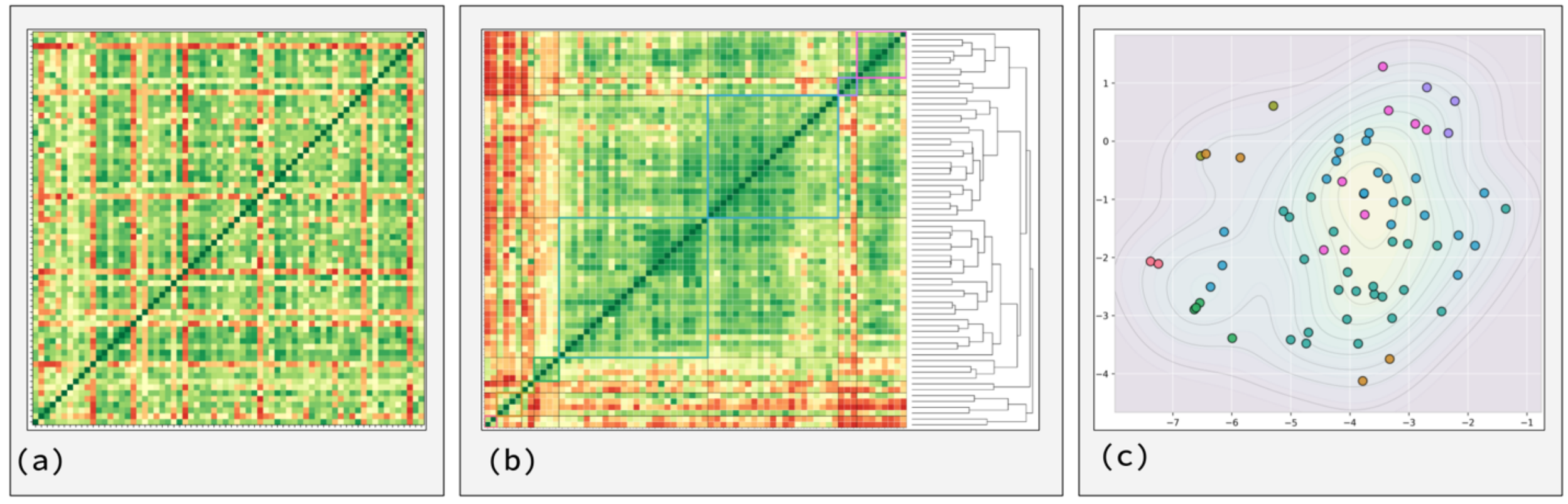

Figure 1: Data transformation process from pairwise correlations to clustered spatial projection. Panel (a) shows the initial pairwise correlation matrix for all 68 definitions, with green indicating high agreement (correlation approaching 1.0) and red indicating disagreement (correlation approaching -1.0). Panel (b) illustrates the sorted correlation matrix, produced after applying hierarchical clustering, with the resulting dendrogram displayed on the right showing the nested relationships between definitions. Panel (c) presents the final t-SNE projection that transforms these high-dimensional correlation patterns into an interpretable 2D semantic landscape, with colors indicating cluster membership and contour lines revealing density variations across the definitional space.

To validate the robustness of our semantic analysis and mitigate potential model-specific biases, we independently replicated all analytical processes across three state-of-the-art LLMs: Claude 3.7 Sonnet, GPT-4o, and Llama-3.3 70B Instruct. These three matrices were then averaged together to produce the final correlation matrix, accounting for cross-model biases and integrating the nuanced perspective of each model in the final feature set. This methodology was implemented using Python, with all code available as an open-source Github repository in the Supplemental Code.

Collection of Expert Definitions and Manual Classification

Here, we were specifically focused on modern researchers' definitions of life (not classic historical figures' definitions). Individuals were selected based on their peer-reviewed papers in the literature which impacted (or required an opinion on) the question of life. These academic authors were contacted through the interview method of email-based correspondence to the interviewee's institutional email, with balanced representation from the fields and disciplines including branches of biology, computer science, physics, and engineering. Sixty-eight total definitions were included in the final analysis. Fourteen distinct countries were represented by the current affiliations of respondents, with the United States (41) and the United Kingdom (9) being the most prevalent.

The respondents were not aware of each other's responses or thought processes, were not given any specific guidance or criteria for inclusion/exclusion and were independently responsible for submitting their own definition. Respondents did not have any set time limit to formulate and structure their responses and were permitted to consult external references and no policy prohibiting contact with colleagues on the matter was set forth. Respondents were instructed to aim for no more than 3 sentences and told that they should think about life as broadly as they wished, not restricted for example to terrestrial life. They were not given any guidelines about word count, and they were made aware that their responses were going to be analyzed in a manuscript for publication.

Two of the provided definitions came from authors of this analysis. Like all the other respondents, the two authors who provided definitions did not see any of the other life definitions submitted, prior to formulating their own. No additional weight was given to these two responses in the downstream analysis.

For editorial purposes, a small proportion of responses were shortened without interfering with the key takeaways of the authors' definitions. This was done to maintain standardized length for all definitions. Finally, experts were given a chance to proof-read their definitions for correct attribution but were strongly discouraged from making any changes at this point, as they had already seen some of the other definitions.

Once all definitions had been received, a process of manual review and categorization was done to ground the results in human understanding before proceeding to the AI-driven analysis. It was possible for an author's definition to be grouped into more than one category. The row "Other" was introduced when a definition did not neatly fit into one of the pre-existing categories. The resulting categorical distinctions are presented in Table 1. All 68 provided definitions of life are presented in Supplemental Table 1.

Quantitative Pairwise Correlation Analysis

To quantify semantic relationships between definitions, we developed a pairwise correlation analysis framework using LLM-based inference. For each pair of definitions in our dataset, we computed a correlation score ranging from -1.0 (complete disagreement) to 1.0 (complete agreement) using the following prompt structure:

```prompt
```

```
Analyze the following two definitions of life, where:
      -1.0 = Fundamentally opposing or incompatible primary frameworks;
      -0.5 to -0.9 = Significantly different emphasis with some
      contradiction;
      0.0 = Independent or orthogonal frameworks;
      0.1 to 0.4 = Slight overlap in secondary elements;
      0.5 to 0.9 = Significant overlap with some differences;
      1.0 = Aligned core frameworks and secondary elements.

Definition 1: {{definition_1}}
Definition 2: {{definition_2}}

What is the correlation metric between -1.0 and 1.0 for these two
definitions?
Respond with ONLY a single number!
```

To enhance reliability and account for potential variability in LLM outputs, we performed multiple replicate inferences (n=3) for each definition pair and computed the average correlation score and standard deviation. This correlation metric was computed for each pair of responses ($n^2$), resulting in a correlation matrix of respondent definitions encoding pairwise similarity ($0 < r < 1$) and dissimilarity ($-1 < r < 0$).

The pairwise correlation analysis generated two n×n matrices, where n is the number of definitions: (1) a correlation matrix M containing the average correlation scores between all definition pairs, and (2) a standard deviation matrix S capturing the variability of correlation estimates. Both matrices were symmetrized by averaging with their transpose ($M' = (M + M^T)/2$) to account for potential rank-ordering effects in the LLM-derived correlation inference, resulting in a total of 6 instances of inference averaged to generate each pairwise correlation metric. The resulting correlation matrices were visualized as heatmaps using a divergent color scheme with correlation values ranging from -1 (disagreement, red) to 1 (agreement, green).

This analysis was repeated with Claude 3.7 Sonnet, Llama 3.3 70B, and GPT-4o, with the final correlation matrix being derived from the average of each analysis' result. Each provided correlation metric is then the result of 18 averaged LLM-generated correlation metrics, 6 from each tested model.

<u>Agglomerative Clustering of Correlation Matrix</u>

To identify natural groupings within the definition space, we applied hierarchical agglomerative clustering to the symmetrized correlation matrix. This bottom-up clustering approach begins with each definition as its own cluster and iteratively merges the most similar clusters until all definitions belong to a single cluster, creating a hierarchical structure that reveals multi-scale relationships between definitions.

The correlation matrix was first transformed into a distance matrix using the relationship $d = \sqrt{2(1-r)}$, where r represents the correlation coefficient. This transformation maps correlation coefficients between -1 and 1 to distances between 0 and 2, with higher correlations corresponding to smaller distances between definitions, with the squaring operation amplifying the distances between high correlations and therefore making the clustering algorithm more sensitive to subtle semantic distinctions between closely related definitions.

We employed complete linkage (maximum distance between any two elements from different clusters, also known as farthest-neighbor linkage) as our agglomerative method. Complete linkage was selected for its ability to produce compact, clearly separated clusters that emphasize conceptual boundaries between semantic groupings, making it particularly suitable for analyzing definitional relationships where clear delineations are desirable.

The dendrogram construction process mimics how one might organize ideas into increasingly broad conceptual categories. At the beginning, each definition stands alone as a unique perspective; each leaf is embedded within a tree with one node. The algorithm then identifies the two most similar definitions and joins them at a height proportional to their similarity. Closely aligned definitions join near the bottom of the dendrogram, while more distant conceptual relatives join higher up. This process continues iteratively, with either individual definitions joining existing groups or groups merging with other groups, always connecting the most similar entities at each step. The resulting tree structure then captures which definitions belong together as related conceptual definitions, as well as the relative semantic distance between them.

To then perform unsupervised clustering on this dendrogram, we analyzed the pattern of merge distances to identify the inflection points where further cluster merges would suddenly bring together substantively different conceptual frameworks. This “elbow” in the derived linkage matrix reveals the natural number of clusters present in the data as the point where merging clusters would begin to obscure important conceptual distinctions. This clustering solution partitions definitions into groups that share fundamental conceptual frameworks while maintaining meaningful separation between distinct philosophical or scientific approaches to defining life. When superimposed on top of the sorted correlation matrix of respondent definitions, this multi-scale perspective allows us to examine both fine-grained distinctions within conceptual frameworks and broader patterns across the entire definitional space.

<u>LLM Cluster Semantic Analysis</u>

For each cluster identified through hierarchical agglomerative clustering, we employed LLMs to perform two complementary analytical processes: (1) intra-cluster thematic analysis to characterize conceptual frameworks and (2) consensus definition generation to distill essential shared elements. Claude 3.7 Sonnet was used to generate the semantic analysis for the multi-model clustered correlation matrix.

The intra-cluster thematic analysis was conducted using a structured prompt designed to systematically extract patterns from definition clusters. This protocol instructed the LLM to analyze definitions through three progressive lenses:

````
```prompt
1. WHAT Are The Core Ideas?
  - List every key concept mentioned
  - Count how often each appears
  - Group similar concepts together
  - Note which concepts appear most
  - Mark which are always present
````
````

```
2. HOW Do Ideas Connect?
  - Find concepts that link to others
  - Map which ideas depend on others
  - Note concept hierarchies
  - Identify central hub concepts
  - Track idea flow patterns

3. WHY This Structure?
  - Identify shared a priori frameworks
  - Find similar starting points
  - Track reasoning patterns
  - Mark scope boundaries
  - Note definitional strategies
```

All prompts are available in their complete form in the Supplemental Code. This thematic analysis was then used to inform our meta-review of various definitions for “what is life?” from a quantitative, statistically grounded, computational point of view. Complementing the thematic analysis, we further employed LLMs to synthesize a single representative definition for each cluster. This protocol explicitly constrained the LLM to adhere to the requirements defined in the following prompt structure:

```prompt
You are tasked with synthesizing a consensus definition of life from a group of experts who have similar perspectives. Below are their definitions:
{{definitions}}

In addition to these provided cluster definitions, you have also conducted a prior thematic review which contains a meta-analysis of the key concepts being discussed by this group of respondents:
{{cluster_analysis}}

- Start: "Life is..."
- Content: Only majority-shared concepts (>50% frequency)
- Language: Technical terms from source definitions
- Structure: Logical flow of connected concepts
- Style: Match source complexity and tone
- Length: Within ±20% of median definition length
- Exclude: Unique views, novel terms, explanations
```

This algorithmic approach to consensus formation ensured that the resulting definitions genuinely represented the central tendencies within each cluster rather than introducing new concepts or arbitrary interpretations. By limiting content to majority-shared concepts and using only terminology present in the source definitions, we maintained fidelity to the original expert perspectives while distilling their shared conceptual core.

The resulting thematic analyses and consensus definitions provide complementary perspectives on each cluster: the thematic analysis characterizes the conceptual landscape and intellectual approach shared by definitions within a cluster, while the consensus definition distills these shared elements into a single coherent statement that represents the cluster's central perspective on what constitutes life. Together, these analyses reveal both the distinctive conceptual frameworks that characterize different approaches to defining life and the core elements that unify definitions within each framework.

Finally, a third inference call was executed to generate a title name for each cluster. Provided with the consensus definition and the thematic analysis in the prompt context, the LLM was asked to generate a brief cluster title according to the following prompt:

```prompt
Based on the consensus definition and thematic analysis provided, generate a SINGLE WORD
or VERY SHORT PHRASE (2-4 words maximum) that captures the DISTINCTIVE ESSENCE of this specific cluster.

The title should be:
1. HIGHLY SPECIFIC to the philosophical, scientific, or conceptual framework unique to this cluster
2. DISTINCTIVE enough that it wouldn't apply equally well to other clusters of life definitions
3. TECHNICALLY PRECISE, using domain-specific terminology where appropriate
4. CONCEPTUALLY FOCUSED on the core unifying principle of these definitions

---

Provided below is the analysis of the given cluster:
{{cluster_analysis}}

Provided below is the derived consensus definition for the given cluster:
{{consensus_definition}}
```

These titles were then mapped to the cluster groups derived by agglomerative clustering and used to define the clusters subsequently plotted by the described 2-dimensional projection methodology.

2-Dimensional Projection of Definitional Landscape

To create an interpretable visualization of the definitional landscape, we applied t-SNE (t-distributed Stochastic Neighbor Embedding) dimensionality reduction to project the high-dimensional correlation features into a two-dimensional representational space. Unlike principal component analysis, t-SNE specifically emphasizes the preservation of local neighborhood relationships, making it particularly suitable for visualizing clusters of semantically related definitions. This was validated by testing against multidimensional scaling (MDS), Principal Component Analysis (PCA), and Uniform Manifold

Approximation and Projection (UMAP), with t-SNE consistently producing the best spatial projection of the underlying correlation embeddings.

The correlation matrix was first transformed into a distance matrix using the relationship $d = \sqrt{(2(1-r))}$, where r represents the correlation coefficient between definition pairs – the same distance transformation applied to agglomerative clustering. The resulting distance matrix was then symmetrized by averaging with its transpose ($M' = (M + M^T)/2$) to account for potential asymmetries in the process of translating correlation features to distance features, effectively placing definitionally similar responses (high correlation) in proximity while separating dissimilar ones (low correlation).

The final t-SNE projection visualizes the semantic landscape of life definitions, with each point representing an expert's response colored according to its assigned cluster. Contour lines were added to highlight density variations within the definitional space, revealing regions of conceptual convergence and divergence. The final visualization incorporates cluster assignments derived from agglomerative hierarchical clustering, with each cluster labeled according to its LLM-derived title that characterizes the unifying conceptual framework of that group. This approach reveals the distinct conceptual territories in relation to the broader definitional landscape, providing a comprehensive spatial reflection of how experts conceptualize life across disciplinary boundaries.

## 4. Results

Expert Definitions

Supplemental Table 1 shows the definitions provided by the survey participants. Some focused on the impossibility (or futility) of a definition, while others attempted this task from varying conceptual alignments. The 68 definitions collected represent a broad spectrum of disciplinary perspectives, from traditional biological frameworks to computational approaches, from physics-inspired thermodynamic views to cognitive and philosophical stances. Despite the likely impossibility of defining truly orthogonal categories along which to categorize or classify these definitions, we attempted to do so manually (without AI-assistance) using the criteria in Table 1.

Manual categorization of these definitions revealed several predominant themes, with certain concepts appearing repeatedly across multiple definitional frameworks. Thermodynamics and energetics appeared in 18 out of 68 (26%) of all definitions, while information/pattern-based approaches were found in 25% of definitions. Dynamics (including self-organization) was the most prevalent theme, appearing in 29% of definitions.

**Table 1**: Categorizing definitions of life

| Criteria: life is defined by specific | Definitions Involving those Criteria |
|---|---|
| Thermodynamics/energetics | Ball, Davies, Dussutour, Froese, Georgiev, Gilbert, Huang, Ingber, Lane, Mitchell, Newman, Picard, Solms, Soto, Szathmáry, Tuszynski, Vallverdú, Nunn |
| Information/pattern | Ackley, Adami, Bongard, Caves, Das, Dodig-Crnkovic, Georgiev, Ingber, Levin, Miller, Mitchell, Gentili, Nunn, Pavlic, Reber, Solé, Witkowski |
| Complexity | Georgiev, Jackson, Sloman, Nunn |
| Computation | Agüera y Arcas, Stepney, Ackley, Witkowski, Dodig-Crnkovic |
| Dynamics (including self-organization) | Agüera y Arcas, Bongard, Brash, Caves, Das, Dodig-Crnkovic, Dussutour, Friston, Froese, Georgiev, Gilbert, Heylighen, Huang, Kauffman, Mitchell, Pizzi, Rechavi, Solé, Szathmáry, Stepney, Witkowski |
| Autonomy/agency/goal-directedness | Ball, Dussutour, Ellis, Jablonka, Calvo, Heylighen, Krakauer, Lander, Lyon, Mitchell, Noble, Shapiro, Solms, Soto, Tuszynski, Vallverdú, Caves |
| Cognition/intelligence | Dussutour, Fontana, Gentili, Krakauer, Levin, Marshall, Miller, Nunn, Picard, Reber, Lyon, Dodig-Crnkovic |
| Structure/architecture | Baluška, Georgiev, Newman, Caves |
| Functionality/behavior | Albantakis, Ball, Caves, Dussutour, Das, Lyon, Sonnenschein, Sultan |
| Replication/Reproduction/ Evolution/Heredity | Ackley, Adamatzky, Adami, Davies, Ingber, Jablonka, Jackson, Lander, Gentili, Nunn, Ratcliff, Shapiro, Sloman, Szathmáry |
| Material composition | Baluška, Das, Kauffman, Lane, Newman, Pizzi, Ratcliff, Sloman |
| Against strong definition | Albantakis, Ball, Frank, Gunawardena, McShea, Stanley, Wong, McShea |
| Other | Ciaunica, Fields, Hoffman, Marshall, Witkowski, Watson |

We further characterized the most prominent divergent properties commonly discussed (Figure 2), revealing that 57% of definitions took an objective stance (defining life in observer-independent terms), while 43% incorporated observer-relative elements. Additionally, 54% of definitions framed life as a continuous property rather than a binary state, reflecting a shift away from categorical distinctions between living and non-living systems. Most definitions (88%) were actionable in the sense that they provided concrete, conventionally defined criteria, while 12% took more inspirational, poetic, or self-recursive approaches.

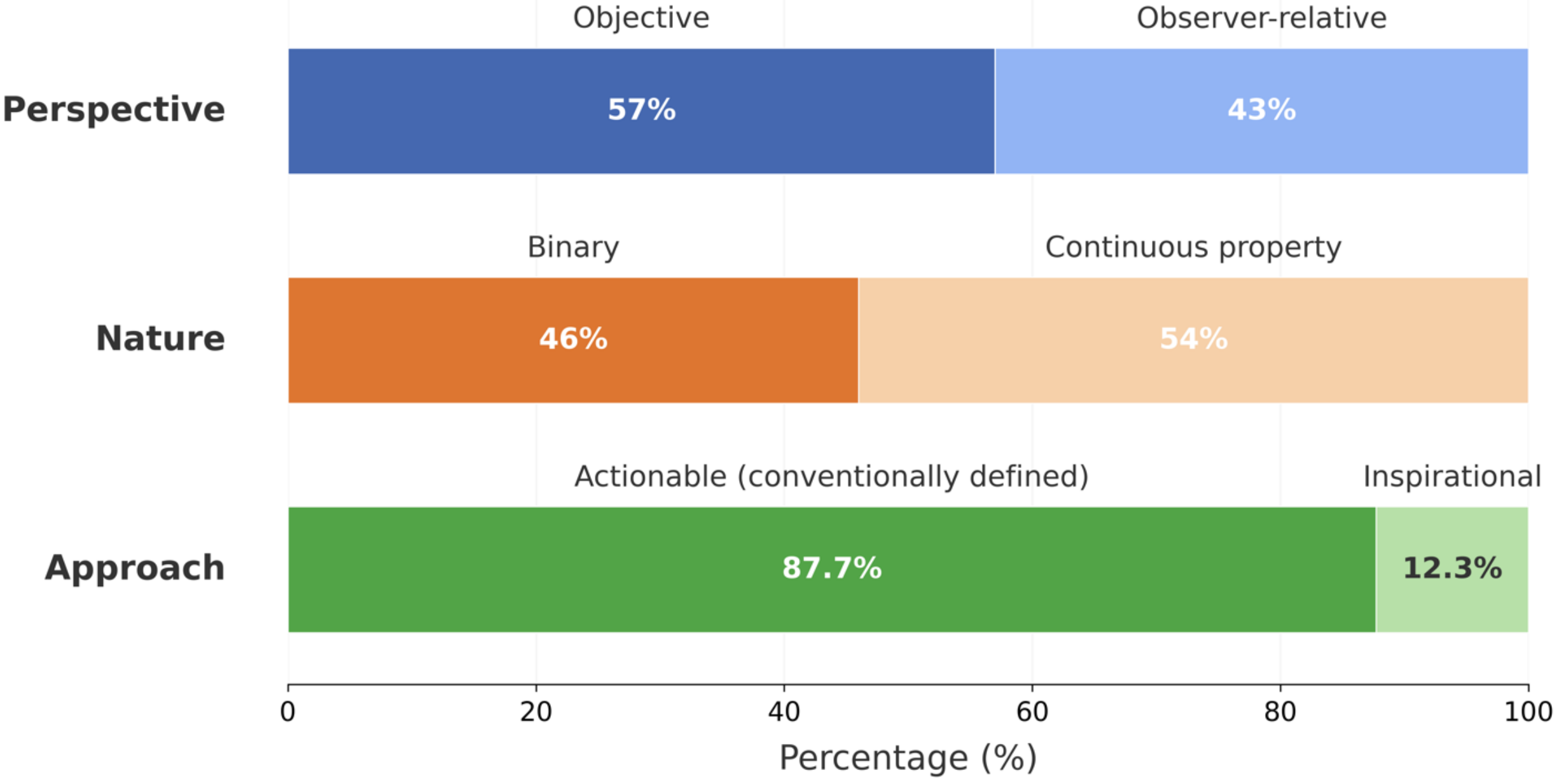

Figure 2: Distribution of general properties among expert definitions of life (n=68).

Pairwise Correlation Analysis

To assess the semantic relationships between expert definitions of life, we analyzed how three state-of-the-art LLMs perceived conceptual overlap among the 68 definitions. Each model independently evaluated all possible definition pairs, producing a correlation matrix that quantifies similarity (-1.0 for complete opposition to 1.0 for perfect alignment) between each definition pair.

Claude 3.7 Sonnet, Llama 3.3 70B Instruct, and GPT-4o were independently used to generate three distinct correlation matrices. Claude 3.7 Sonnet demonstrated the most optimistic assessment with the highest average correlation, while Llama 3.3 70B Instruct produced more conservative estimates of correlation and showed greater standard deviation in its responses. Notably, all models displayed negative skewness (-0.76 to -0.97), indicating a systematic tendency to identify more areas of conceptual similarity than opposition. The general correlation matrix statistics for each model are presented in Supplemental Table 2.

Despite the models' independent analysis, they showed remarkable consistency in their overall assessment of which definitions aligned conceptually and which diverged. The pairwise matrix-to-matrix correlations were notably high:

- Claude 3.7 Sonnet vs. Llama 3.3 70B Instruct: r = 0.7279
- Claude 3.7 Sonnet vs. GPT-4o: r = 0.8103
- Llama 3.3 70B Instruct vs. GPT-4o: r = 0.7977

These substantial correlations suggest that while each model brought a distinct analytical perspective to the definitional landscape, they independently converged on similar patterns of conceptual relationships.

These correlation matrices are shared in Supplemental Figures 1, 2, and 3 for Claude, Llama, and GPT respectively. The resulting unsorted multi-model averaged correlation matrix is presented as Supplemental Figure 4. Additionally, the standard deviations (Entropy Matrix) for all aggregated instances of inference are available at Supplemental Figures 5, 6, and 7 for Claude, Llama, and GPT respectively, highlighting the consistency with which the LLMs scored correlation for each respondent across repeated instances of inference.

To mitigate potential biases from any single model and create a more robust assessment of semantic relationships, we computed an element-wise average across the three symmetrized correlation matrices. The multi-model integration succeeded in capturing the central tendencies of all three individual analyses, as evidenced by the high correlations between each model's matrix and the averaged matrix:

- Claude 3.7 Sonnet vs. Multi-Model-Average: r = 0.9017
- Llama 3.3 70B Instruct vs. Multi-Model-Average: r = 0.9294
- GPT-4o vs. Multi-Model-Average: r = 0.9358

The multi-model average matrix successfully captured the nuanced perspectives of each model, ultimately producing a rich encoding of pairwise correlation patterns amongst the provided definitions of life.

Agglomerative Clustering of Correlation Matrices

The three LLMs tested exhibited distinct clustering behaviors, identifying between 7 and 11 natural clusters through elbow method analysis. Claude 3.7 Sonnet produced the most integrative solution with 7 clusters and the lowest contrast metric (0.258), while Llama 3.3 70B Instruct demonstrated the most discriminative approach with 11 clusters and the highest contrast metric (0.563). The multi-model integration yielded 8 clusters with a balanced contrast metric of 0.359, positioning it between the integrative and discriminative extremes. This integrated approach achieved robust intra-cluster correlation (0.509 ± 0.187) while maintaining modest inter-cluster correlation (0.150 ± 0.390), suggesting it captured both the distinctiveness of conceptual frameworks and meaningful connections between them. The statistics of each tested model's clustering results are available in Supplemental Table 3.

Figure 3 presents the visualization of the multi-model integrated analysis, showing both the clustered correlation matrix (with green indicating high correlation and red indicating negative correlation) and the corresponding dendrogram structure on the right. Additionally, the clustered correlation matrices for Claude, Llama, and GPT's independent analyses are presented in Supplemental Figures 8, 9, and 10 respectively to show how each model uniquely organized the respondents in relation to each other.

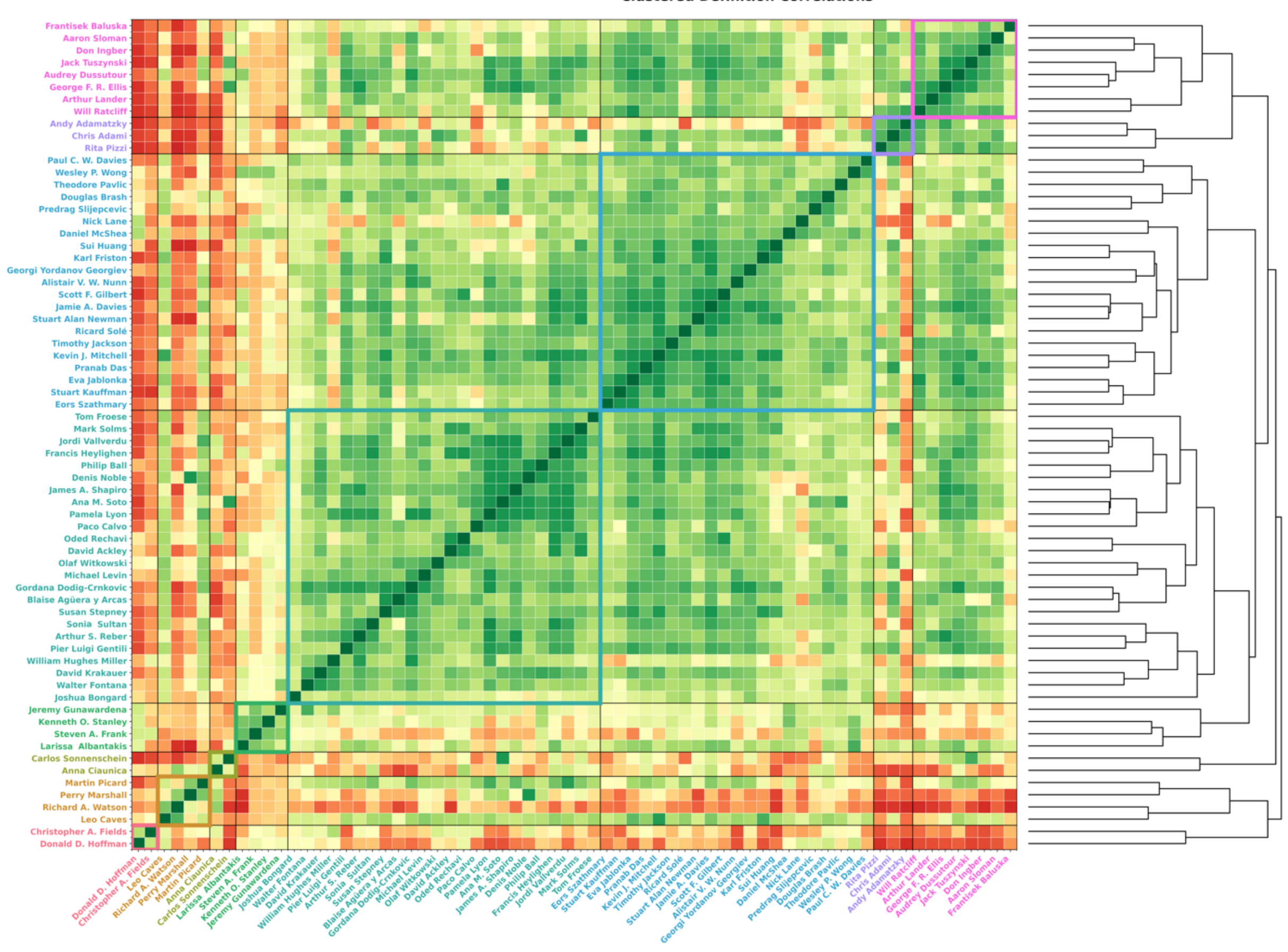

Figure 3: Hierarchically sorted correlation matrix of respondent definitions, along with the corresponding dendrogram defining the linkage distance between definitions, highlighting 8 distinct conceptual clusters.

This sorted correlation matrix reveals a distinct block-diagonal structure indicating strong intra-cluster coherence, while also exhibiting meaningful cross-cluster correlations that suggest conceptual continuity across the definitional landscape. To investigate how consistently the various LLMs grouped the respondents into the same cluster relationships, we compared the clustering consistency from each of the three models' analyses plus the multi-model integration.

- Claude 3.7 Sonnet and Llama 3.3 70B Instruct: 71.2% consistency
- Claude 3.7 Sonnet and GPT-4o: 74.8% consistency
- Claude 3.7 Sonnet and Multi-Model-Average: 74.1% consistency
- Llama 3.3 70B Instruct and GPT-4o: 75.7% consistency
- Llama 3.3 70B Instruct and Multi-Model-Average: 72.8% consistency
- GPT-4o and Multi-Model-Average: 79.1% consistency

These high cross-model consistencies (71.2% to 79.1%) indicate substantial agreement on which definitions should be grouped together, despite the different number of clusters identified by each model.

To further assess the robustness of our clustering, we examined how consistently individual definitions were grouped across the three models. The overall grouping stability was 53.7%, indicating that just over half of all definition pairs were consistently grouped together (or consistently separated) across all models. This moderate stability suggests that while the broad structure of the definitional landscape is robust, there remains genuine ambiguity in how some definitions should be categorized at the boundaries between clusters.

Some definitions showed remarkably high grouping consistency across models: those by Christopher Fields, Donald Hoffman (both 100% consistency), Andy Adamatzky, Leo Caves, and Richard Watson (all 95.5% consistency). These definitions appear to occupy clear, distinctive positions in the semantic landscape. In contrast, definitions by Mark Solms (28.4%), Aaron Sloman (26.9%), Paul Davies (23.9%), Tom Froese (23.9%), and Wesley Wong (22.4%) showed the lowest consistency, suggesting these definitions span multiple conceptual frameworks or occupy boundary positions between established clusters. Looking at Figure 3, we can observe that stable definitions tend to appear within the most distinctly colored blocks, while unstable ones often appear in transition zones between clusters.

Intra-Cluster Semantic Analysis

From the 8 distinct clusters derived by agglomerative clustering of the averaged multi-model pairwise correlation matrix, the thematic content of each cluster was further explored. The consensus definitions in Table 2 represent distillations of the core concepts shared within each cluster, providing insight into how different perspectives approach the question of what constitutes life. These definitions were derived through LLM analysis of the definitions within each cluster, extracting the majority-shared concepts while maintaining the technical language and conceptual frameworks characteristic of each group.

**Table 2:** Cluster Titles and Consensus Definitions

| Title | Consensus Definition |
|---|---|
| Perceptual Categorization | *Life is a perceptual category arising from how systems that consider themselves alive recognize and categorize others, rather than an objective distinction with intrinsic properties. The apparent boundary between living and nonliving exists primarily as an artifact of our sensory limitations that introduce artificial distinctions into what may be a more unified reality.* |
| Self-Sustaining Dynamic Patterns | *Life is a persistent dynamic pattern that sustains itself through recursive processes, creating and maintaining its own conditions for existence. It manifests as a self-aware system capable of transformation, emerging through the resonance and harmonic interaction of its components. This pattern exists in relationship with its environment, expressing consciousness while continuously recreating itself through mutually transformative connections.* |
| Dynamic Relational Process | *Life is a permanent movement characterized by dynamic exchange between internal and external environments, fundamentally requiring relationship with pre-existing others. Life cannot exist in isolation or stasis, as its essential nature involves both movement and proliferation. Rather than being something humans can create, life is received and passed on, existing as an ongoing process of interconnection.* |
| Pragmatic Definitional Skepticism | *Life is a contextual construct better approached through functional utility than universal definition, where definitional efforts should serve specific research purposes rather than establish absolute boundaries. The concept exists at the intersection of scientific and philosophical domains, requiring recognition of disciplinary limitations and pragmatic focus on what advances understanding rather than what constitutes a "correct" characterization.* |
| Cognitive Autonomy | *Life is a self-maintaining, goal-directed system that processes information to adapt to environmental changes while preserving its organizational boundaries. It operates as an autonomous agent that actively opposes entropy through energy exchange, making purposeful decisions that ensure its continued existence. Life functions as a cognitive process that senses, interprets, and responds to its surroundings, modifying itself when necessary to maintain stability despite changing conditions.* |
| Dissipative Self-Organizing Systems | *Life is a dynamic, far-from-equilibrium process characterized by self-organization and self-maintenance through controlled energy dissipation. It maintains semi-permeable boundaries that separate the system from its environment while allowing selective exchange of matter and energy. Living systems process, store, and transmit information through feedback mechanisms that enable adaptation and evolution over time.* |
| Informational Self-Replication | *Life is a self-replicating system of information encoded in a physical substrate, capable of reproducing itself while maintaining order against natural decay. This information-based process enables the transmission of complex structural and functional data across generations, with the physical components serving primarily as carriers for this essential informational content.* |
| Self-Replicating Thermodynamic Systems | *Life is a self-sustaining physical and chemical system that reproduces itself while maintaining organization away from thermodynamic equilibrium. It interacts adaptively with its environment, absorbing and transforming resources to support its replication and growth. Living systems exhibit autonomy through self-regulation and response to stimuli, while their reproductive capabilities enable evolutionary processes.* |

The complete thematic analysis performed for Claude, Llama, GPT, and the resulting multi-model average are shared in Supplemental Files 1, 2, 3, and 4 respectively.

t-SNE Projection of Life’s Definitions

t-SNE dimensionality reduction transformed the high-dimensional correlation data into an interpretable two-dimensional visualization of the 68 expert definitions (Figure 4). t-SNE was chosen for its ability to preserve local neighborhood structure while revealing global patterns in high-dimensional data, but the complete panel of clustering results from Multidimensional Scaling (MDS), Uniform Manifold Approximation and Projection (UMAP), t-SNE, and an ensemble average is presented in Supplemental Figure 11.

The resulting t-SNE projection of Figure 4 reveals a continuous semantic landscape organized along two conceptually interpretable axes. Dimension 1 (horizontal) maps the transition from observer-dependent, perceptual frameworks (left) to objective, material-structural definitions (right). The leftmost position is occupied by the isolated Perceptual Categorization cluster, exemplified by Fields' recursive formulation: "to be alive is to be considered alive by systems that consider themselves alive." This contrasts sharply with the right-side clusters that emphasize concrete mechanisms like replication and thermodynamic properties, captured by Adami's conception of life as "information that can replicate itself."

Dimension 2 (vertical) reveals a fundamental ontological tension between entity-based and process-based conceptualizations of life, recapitulating the historical dialectic from Cartesian mechanism to Aristotelian teleology. The topmost positions are occupied by definitions emphasizing material structures, component properties, and tangible mechanisms, such as Baluška's "Living Cells and All Their Constructs," Pizzi's focus on "carbon-based elements with self-organizing and self-replicating properties," and Sloman’s "components that are able to extend and replicate themselves.” The bottom region contains process-oriented, cognitive-phenomenological frameworks, represented by Marshall's metaphorical "fire that lights itself," Picard's "self-sustaining system capable of healing that uses and organizes energy to achieve expression(s) of consciousness," and Noble’s “self-creating agency.” This vertical gradient captures the perennial tension between substantialist perspectives that view life as a collection of specified entities (echoing Cartesian and Hobbesian frameworks) and teleological views that conceptualize life as coherent patterns of organization, information flow, and emergent self-actualization.

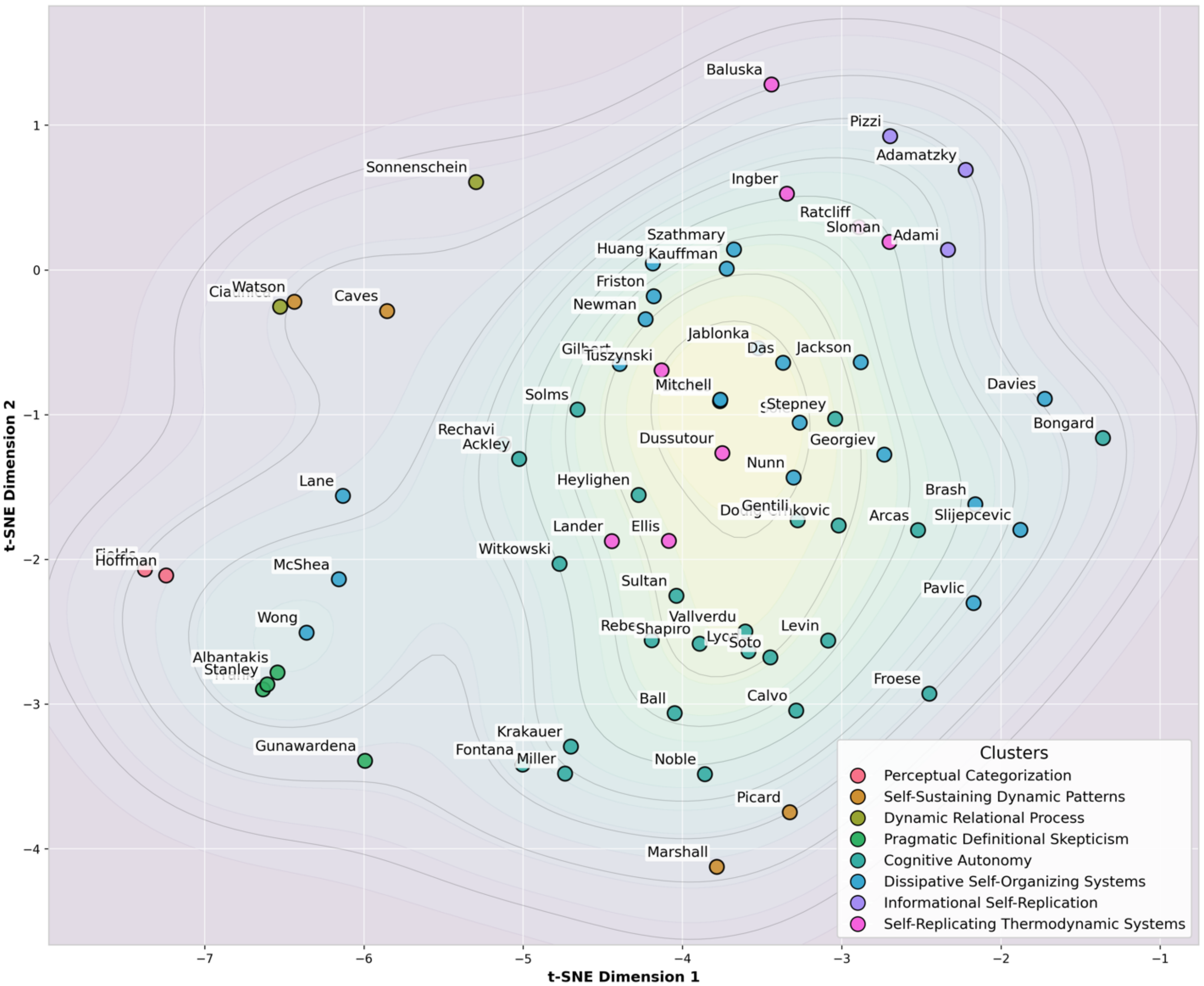

Figure 4: Two-dimensional t-SNE definitional landscape showing the distribution of expert perspectives across eight conceptual clusters, derived from the average pairwise correlation matrices from independent Claude 3.7 Sonnet, Llama 3.3 70B Instruct, and GPT-4o analyses.

The dominant central attractor comprises two partially overlapping clusters: Cognitive Autonomy (teal, n=24) and Dissipative Self-Organizing Systems (blue, n=21). These concentrate 66% of definitions in a high-density zone where physical, informational, and agential perspectives are integrated by a variety of definitional approaches.

Contour patterns expose varying semantic densities across the landscape. The sparsest regions surround the isolated Perceptual Categorization cluster, while gradient transitions appear between process-oriented clusters (orange and yellow) and the central attractors. The right quadrant displays two distinct poles: Informational Self-Replication (purple, n=3) achieves maximum structural distance from perceptual clusters, while Self-Replicating Thermodynamic Systems (pink, n=8) bridges traditional biological perspectives with information-theoretic approaches.

The continuous nature of this semantic space is further illuminated through definitions that function as conceptual bridges between cluster territories. Watson's characterization of life as "the pattern and process of love — a deeply vulnerable mutual dance" creates a semantic pathway between perceptual-relational and teleological frameworks. Similarly, Noble's "Life is self-creating agency" occupies a transitional position between process-oriented approaches and cognitive frameworks, explicitly connecting Aristotelian self-actualization with modern agency-based perspectives. Levin's "Living beings remember, and anticipate, coarse-graining experience into a continuous actionable dream" bridges cognitive and physical perspectives through concepts that unite information processing with thermodynamic constraints, demonstrating how contemporary definitions continue to negotiate the historical tension between mechanism and teleology.

The emergence of interpretable dimensions from computational semantic analysis validates this approach as a methodological bridge between reductionist and holistic perspectives on fundamental questions in science and philosophy. Rather than imposing preconceived categories, the methodology reveals latent structure within expert discourse, demonstrating how computational tools can expose patterns of conceptual coherence that might otherwise remain obscured by disciplinary fragmentation. The high convergence across different LLMs (correlation coefficients >0.7) further validates the robustness of these patterns, while the consistent identification of conceptual bridges and transitional zones across models suggests these represent genuine features of the definitional landscape rather than analytical artifacts.

## 5. Discussion

Cluster Semantic Analysis

The eight conceptual clusters revealed through agglomerative clustering present distinct yet interconnected perspectives within the definitional landscape, each occupying a unique semantic position that illuminates fundamental tensions and convergences in contemporary understandings of life.

The Perceptual Categorization cluster (n=2) occupies the most divergent semantic space, achieving maximum conceptual distance from material-structural frameworks. Fields' recursive formulation — "to be alive is to be considered alive by systems that consider themselves alive" — challenges foundational assumptions about objective reality. Hoffman's positioning of the living/nonliving boundary as "an artifact of the limitations of our senses" resonates with quantum mechanical insights about observer effects, suggesting unexplored connections between life's definition and fundamental physics.

Self-Sustaining Dynamic Patterns (n=4) bridges radical epistemology with process philosophy through sustained focus on relationship and consciousness. Watson's characterization of life as "pattern and process of love — a deeply vulnerable mutual dance" introduces affective dimensions absent from mechanistic frameworks, while Ciaunica's insistence that life "cannot exist without an other, already being there, already alive" poses significant challenges to origin-of-life theories that assume primordial isolation. This cluster's correlation patterns reveal conceptual affinity with both observer-dependent frameworks and certain elements of Cognitive Autonomy, suggesting potential integration pathways between phenomenological and scientific approaches.

Dynamic Relational Process (n=2) distills relationality to its kinetic essence, with Sonnenschein's declaration that "Without proliferation and movement, there is no life" establishing clear boundary conditions. This minimal cluster creates bridges between abstract relationality and classical biological requirements, positioning itself as a phase-shift between philosophical and empirical approaches. Its adjacent positioning to Self-Sustaining Dynamic Patterns reflects deep conceptual affinity while its moderate positive correlations with replication-focused clusters suggest this bridge function operates effectively across multiple scales.

Pragmatic Definitional Skepticism (n=4) functions as methodological counterpoint, with Lane's assertion that "drawing a line across a continuum is always arbitrary" challenging the entire enterprise of categorical definition. This cluster's unique correlation pattern (negative relationships with strong boundary conditions, positive relationships with multidimensional frameworks) reveals it functions as conceptual mediator between competing definitional paradigms rather than advocating for any particular position. Frank's characterization of definitions as "tools and not endpoints" provides philosophical grounding for the computational methodology employed in this study.

Cognitive Autonomy (n=24) emerges as the primary conceptual attractor, its central positioning and high membership reflecting broad cross-disciplinary convergence. The cluster's internal diversity spans cybernetic perspectives ("information pattern capable of self-reproducing") to phenomenological approaches ("self-awareness and self-agency"), revealing how cognitive frameworks accommodate both mechanistic and experiential dimensions. Notable is Dodig-Crnkovic's stark equation "Life = cognition," which frames computational processes as fundamental rather than derivative. The cluster's moderate positive correlations across nearly all other frameworks (excluding Perceptual Categorization) suggests it functions as semantic hub rather than a vanguard position.

Dissipative Self-Organizing Systems (n=21) provides complementary gravitational pull through emphasis on physical principles. Its substantial overlap with Cognitive Autonomy creates the definitional landscape's central convergence zone — a high-density region where 66% of all definitions concentrate. Pavlic's characterization of life as "statistical novelty producer" bridges thermodynamic necessity with evolutionary innovation, while Jackson's emphasis on "unique mechanisms of information storage and retrieval" connects physical processes to functional emergence. The cluster bridges mechanical and functional approaches through explicit focus on energy gradients and boundary conditions that enable rather than merely constrain biological organization.

Informational Self-Replication (n=3) represents the definitional space's reductionist pole, with Adami's formulation — "Life is information that can replicate itself" — providing maximum precision at potential cost of comprehensiveness. This cluster's correlation patterns reveal sharp boundaries that exclude observer-dependent and relational frameworks while maintaining strong positive correlations with Self-Replicating Thermodynamic Systems. The cluster's small size belies its conceptual influence as an anchoring point for information-theoretic approaches, though strict criteria on the medium for informational self-replication could exclude systems such as prions or crystalline life forms that reproduce without genetic information.

Self-Replicating Thermodynamic Systems (n=8) integrates traditional biological requirements with physical constraints, creating a bridge between classical genetics and emerging physics-based frameworks. Dussutour's enumeration of interrelated capacities — reproduction, autonomy, responsiveness, plasticity, energy utilization — exemplifies the multi-property approach that characterizes this cluster. The internal diversity spans from cellular-structural emphasis (Baluška's "Living Cells and All Their Constructs") to process-focused perspectives (Ingber's emphasis on energy conversion), revealing tensions even within traditional biological formulations of life.

The correlation patterns between these clusters reveal a semantic topology where conceptual proximity reflects underlying philosophical affinity. The mathematical structure of inter-cluster relationships exposes distinct patterns where central clusters (Cognitive Autonomy, Dissipative Self-Organizing Systems) exhibit broad positive correlations across the landscape, while peripheral clusters display selective affinity patterns that create conceptual archipelagos rather than isolated islands. This field-like structure, with multiple attractors exerting varying degrees of gravitational pull, challenges simple spectral models of definitional space.

Cluster boundaries function as transitional zones rather than hard demarcations, with definitions at peripheries often exhibiting properties of neighboring frameworks. These semantic gradients facilitate conceptual migration and explain why certain definitions show low grouping stability across LLM analyses — they occupy genuine boundary regions where multiple clustering solutions remain mathematically valid. The persistence of these archetypal themes across different LLM architectures (correlation coefficients >0.7 between models) validates them as robust features of expert discourse rather than imposed taxonomies.

From Categorical Definition to Continuous Semantic Space

Our computational analysis of 68 expert definitions reveals life as a continuous semantic landscape rather than discrete categorical states. The t-SNE projection (Figure 4) demonstrates this continuity through two interpretable dimensions: Dimension 1 (x) maps the gradient from observer-dependent, relational frameworks to objective, material-structural approaches, while Dimension 2 (y) reveals the transition from process-oriented, teleological perspectives to entity-based, structural frameworks.

The ancient vitalist-mechanist dichotomy reemerges along Dimension 1, with vitalist-adjacent perspectives clustering toward the perceptual-relational pole on the left and mechanistic frameworks toward the material-structural pole on the right. Similarly, the historical tension between substance-based and process-based ontologies maps directly onto Dimension 2, revealing that these philosophical divides persist in contemporary scientific discourse.

The t-SNE visualization further reveals patterns in how expert perspectives distribute across this conceptual terrain. The highest density region centers on the overlap between Cognitive Autonomy (n=24) and Dissipative Self-Organizing Systems (n=21) clusters, containing 66% of all definitions. This concentration suggests emerging consensus around frameworks that integrate:

- Physical principles (far-from-equilibrium thermodynamics, energy dissipation)
- Informational processes (boundary maintenance, information storage and transmission)
- Functional capacities (self-organization, environmental interaction)
- Agential properties (goal-directedness, adaptive behavior)

This convergence zone bridges traditionally separate disciplinary approaches, from physics and biology to cognitive science and information theory. The integrative nature of high-density definitions, exemplified by Agüera y Arcas's "self-modifying computational phase of matter arising from evolutionary selection for dynamic stability," suggests that the zeitgeist is moving toward hybrid frameworks that transcend historical dichotomies.

Peripheral regions of the semantic landscape, while less densely populated, may indicate emerging paradigm shifts. The isolated Perceptual Categorization cluster (Fields, Hoffman) challenges the implicit realism underlying most definitional approaches, while the Pragmatic Definitional Skepticism cluster questions the enterprise of universal definition itself. These vanguard positions may gain relevance as scientific advancement increasingly encounters borderline cases between living and non-living systems.

The semantic landscape provides a novel framework for positioning such ambiguous entities. Viruses occupy transitional zones between Informational Self-Replication and Self-Replicating Thermodynamic Systems clusters, while artificial intelligence systems might trace distinctive trajectories through this space as they develop increasingly sophisticated capabilities. Current LLMs might occupy positions between "Cognitive Autonomy" and "Informational Self-Replication" clusters — exhibiting information processing and goal-directed behavior while lacking material autonomy and self-maintenance. Future systems incorporating embodied robotics might migrate toward the "Dissipative Self-Organizing Systems" region as they develop capacities for environmental interaction and physical self-maintenance.

Methodological Innovations and Limitations

It should be noted that we make no claim about these definitions being a statistically representative survey of any specific field. Instead, we collected the opinions of a hand-picked group of scholars to develop methods for analysis of current thought on this topic and similarly difficult areas. We extend our apologies in advance to any collaborators or external experts who were not asked to share their definition of life for this analysis, but our code and workflow make possible future surveys with much larger statistical power to study the thoughts of specific communities.

Our application of LLM-driven semantic analysis to the definitional landscape demonstrates a novel approach to consensus formation across disciplinary boundaries. Rather than seeking to identify a single "correct" definition, this approach maps the conceptual territory within which various definitions operate, revealing both their relationships and distinctive contributions. This method offers a model for addressing other contested definitional territories across the sciences.

This approach differs from traditional aggregation methods by preserving the distinctiveness of competing frameworks while revealing their underlying relationships. Unlike Delphi methods that drive toward consensus through iterative refinement, the presented computational meta-analysis maintains the integrity of diverse perspectives while providing a quantitative map of their interrelationships. This methodology bridges reductionist approaches that seek necessary and sufficient conditions with pluralist approaches that embrace multiple definitions, thus revealing a unified semantic landscape in place of fragmented definitional domains.

Several limitations should be acknowledged. First, the expert definitions analyzed, while diverse, cannot claim to represent all possible perspectives on life. The 68 respondents asked to provide definitions were hand-curated by the authors, and are intended not to serve as a representative sample of the scientific consensus but rather a selected set of cross-disciplinary experts. Second, the pairwise correlation methodology, while rigorous, necessarily relies on the semantic capabilities of the LLMs employed, which may introduce subtle biases in how conceptual relationships are evaluated. In addition to the limitations presented by the LLM's competency, the computational time and cost for the generation of pairwise correlation matrices scale as the square of the number of samples, making large cohort analyses prohibitively expensive without highly permissive rate-limits or self-hosted LLMs. Third, the dimensional reduction required for visualization inevitably sacrifices some nuance in the relationships between definitions.

Despite these limitations, the methodology demonstrates the potential for computational semantic analysis to reveal patterns of conceptual coherence that might otherwise remain obscured by disciplinary fragmentation. This approach suggests that apparent definitional disagreements may, in some cases, reveal deeper patterns of coherence when viewed as gradients in semantic space.

## 7. Conclusion

Our computational analysis of 68 expert definitions presents life as a continuous semantic landscape rather than discrete categorical states. LLM-driven semantic analysis identified two emergent dimensions — observer-dependent versus objective frameworks, and process-based versus entity-based ontologies — that recapitulate ancient philosophical tensions while providing quantitative tools for mapping contemporary expert perspectives. This approach transforms apparent conceptual disagreements into complementary positions within a unified definitional space.

The methodology preserves competing perspectives while revealing quantitative relationships between them, bridging reductionist and pluralist paradigms. Concentration of 66% of definitions within overlapping Cognitive Autonomy and Dissipative Self-Organizing Systems clusters indicates emerging consensus on frameworks integrating physical principles, informational processes, and agential properties. This convergence suggests the field is moving beyond historical dichotomies toward hybrid approaches that transcend disciplinary boundaries.

The diversity of life's definitions reveals not conceptual dissonance but a structured semantic topology with predictable patterns of convergence and divergence. The eight archetypal clusters identified here represent stable conceptual attractors that persist across different computational analyses, suggesting these are robust features of expert thinking rather than analytical artifacts. Critically, the emergence of interpretable dimensions without explicit encoding demonstrates that fundamental philosophical tensions — observer-dependence versus objectivity, process versus substance — continue to structure contemporary scientific discourse in systematic ways.

This methodology scales naturally to larger populations across adjacent fields. While our analysis focused on a curated set of experts, the computational framework enables systematic analysis of thousands of responses across diverse disciplines. Such expansive polling could reveal disciplinary-specific patterns within the broader semantic landscape, tracking how field-specific vocabularies and methodological commitments shape definitional approaches. This larger-scale mapping might expose previously unrecognized connections between fields or identify emerging conceptual territories at disciplinary boundaries.

Beyond life itself, cognitive processes present the next frontier for this analytical approach. Cognition, like life, resists simple definition and spans multiple levels of analysis from neural mechanisms to subjective experience. Memory, with its complex interplay of molecular, cellular, and systemic processes, exemplifies how fundamental concepts in biology traverse similar definitional challenges. Applying our computational semantic framework to these contested territories could reveal whether certain definitional patterns represent general features of complex biological phenomena or unique characteristics of life's conceptualization.

The question "what is life?" persists because it maps a multidimensional conceptual space rather than awaiting a singular answer. Our methodology offers immediate applications to other contested definitions across sciences, particularly as synthetic biology and artificial intelligence challenge traditional boundaries. Future progress on fundamental questions may arise not from establishing rigid demarcations but from understanding how perspectives naturally converge and diverge within definitional spaces.

**Acknowledgements**

Chris Fields would like to acknowledge Eric Dietrich as the inspiration for his definition. We thank all of the contributors for their thoughtful definitions. We also thank Julia Poirier for assistance with the manuscript. Michael Levin gratefully acknowledges support from the Elisabeth Giauque Trust, London, and Grant 62212 from the John Templeton Foundation.

## What Lives? – Supplemental

***Supplemental Code:*** *Github Repository for the complete source code repository for this analysis of "What Lives?"*

https://github.com/mrbende/what-lives/

***Supplemental File 1:*** *Claude 3.7 Cluster Thematic Analysis Results*

https://github.com/mrbende/what-lives/blob/main/supplemental/supplemental_f1_claude37_cluster_analysis.md

***Supplemental File 2:*** *Llama 3.3 70B Instruct Cluster Thematic Analysis Results*

https://github.com/mrbende/what-lives/blob/main/supplemental/supplemental_f2_llama3_cluster_analysis.md

***Supplemental File 3:*** *GPT-4o Cluster Thematic Analysis Results*

https://github.com/mrbende/what-lives/blob/main/supplemental/supplemental_f3_gpt4o_cluster_analysis.md

***Supplemental File 4:*** *Multi-Model Cluster Thematic Analysis Results, Generated with Claude Sonnet 3.7*

https://github.com/mrbende/what-lives/blob/main/supplemental/supplemental_f4_multi_model_cluster_analysis.md

***Supplemental Table 1:*** *Definitions of Life*

| Name | Definition |
|---|---|
| Aaron Sloman | Life exists on a physical/chemical entity (e.g. galaxy, or planet) containing components that are able to extend and replicate themselves using mechanisms that repeatedly add new forms of complexity by absorbing structures from their environment, including repeatedly extending the forms and mechanisms of replication and their powers. |
| Alistair Nunn | Life can be defined as a self-organising adaptive "intelligent" entity based on existing atoms/compounds that combine into an auto-catalytic network driven by thermodynamic dissipation of energy, which maintains its existence through internal natural selection of smaller dissipating components to maintain function, where memory to respond, maintain and adapt to the environment is held in the fields created by the movement of charged entities that through a process of feedback, guide the formation and maintenance of a matter-based system that can maintain these fields enabling long term memory, which defines self from non-self. |
| Ana Soto | A biological system differs from an abiotic system in that the former is a goal-pursuing normative agent, capable of creating its own norms, and thus able to change. At a minimum, being alive is a precarious state in continuous search of nutrients to keep itself alive. Additionally, such a system creates novelty. In our terrestrial world, this is illustrated by the built-in ability of cells to move and proliferate, creating variation at each cell division (default state). |
| Andy Adamatzky | Reproduction. with mutations. |
| Anna Ciaunica | Life is the permanent movement of our bodies towards a balanced exchange between what lays inside and what is outside. Life is what cannot exist without an other, already being there, already alive. Life is not something we can create or make. Life is something we have received, like a gift. |
| Arthur Lander | To me, life refers to those systems that arise when natural selection gives rise to automatic control. Natural selection here is to be understood the way biologists do: adaptive change enabled by reproduction and selective survival. Automatic control here is to understand the way engineers do: mechanisms that achieve robust performance in the face of unpredictable disturbances. When the latter emerges as a result of the former, I call that life. |
| Arthur Reber | Life: An entity that can process ambiguous information in its environment and make adaptive decisions that mitigate stress, promote growth, and encourage reproduction can be said to be alive. |
| Audrey Dussutour | I understand life as dynamic process emerging from networks of interactions between units (self-organization) characterised by the ability to:<br>- reproduce<br>- be autonomous (self-regulation)<br>- respond to stimuli (being external or internal)<br>- be plastic<br>- produce and use energy |
| Blaise Agüera y Arcas | Life is a self-modifying computational phase of matter arising from evolutionary selection for dynamic stability; it complexifies through symbiotic composition. The "matter" doesn't have to be physical matter in our universe, but can be in any environment that supports computation, including of course on a computer. |
| Carlos Sonnenschein | Without proliferation and movement, there is no life. |
| Chris Adami | Life is information that can replicate itself. Life is a property of an ensemble of units that share information coded in a physical substrate and which, in the presence of noise, manages to keep its entropy significantly lower than the maximal entropy of the |

| Name | Definition |
| --- | --- |
| | ensemble, on timescales exceeding the "natural" timescale of the decay of the (information-bearing) substrate by many orders of magnitude. |
| Chris Fields | To be alive is to be considered alive by systems that consider themselves alive. |
| Daniel McShea | 1) Biology is not yet ready to give a formal definition of life, that is, a definition in the philosopher's sense of necessary and sufficient conditions. (Indeed, a good argument can be made that life is not a natural kind, in which case necessary and sufficient conditions will never be found.) 2) When we understand it better, life is likely to turn out to be multidimensional, like "happiness,", not reducible to one or even a small number of independent variables. Things like distance from thermodynamic equilibrium, metabolic closure, developmental and regenerative near-closure, complexity in the sense of number of part types, complexity in the sense of number of levels of nestedness, memory capacity, and many more will all be relevant and will vary somewhat independently from one instance of life to another. 3) Being alive is likely to turn out not to be dichotomous, but to vary continuously, like everything else in the universe, in all dimensions, ranging from the extremely alive to the negligibly alive, with the entire range of intermediate values occupied (albeit not equally densely). Averaged over many dimensions of life-li-ness (life-i-tude-in-osity?), earthworms are probably very alive. Viruses much less so. Hurricanes only a little bit. Rocks negligibly. |
| David Ackley | Life is any machine that works to preserve its patterns. |
| David Krakauer | Autonomous problem-solving matter. |
| Denis Noble | Life is self-creating agency. |
| Don Ingber | Life refers to physical structures that form through hierarchical self-assembly of multiple different types of molecular components that exhibit the ability to harness energy and convert it into new materials, which it then uses to grow and reproduce itself. |
| Donald Hoffman | The distinction between living and nonliving objects is not principled. It is an artifact of the limitations of our senses. We always interact with Life, but our senses serve as interfaces that dumb things down, and thereby introduce an apparent distinction between living and nonliving. |
| Douglas Brash | A process (not a thing, substance, property, or state) so it is better thought of as Living. Living is defined as a process of development, building higher hierarchical levels out of lower levels. |
| Eörs Szathmáry | A living world unfolds in a population of evolutionary units. Living systems are ontologically rooted in autocatalytic chemistry in which metabolism, boundary and a control system are acting together to form a Kantian whole. Individual living systems are not necessarily units of evolution, and a living world can exist without living systems, although the overlap between these sets is huge. |
| Eva Jablonka | Life, which is enduring enough to be found or constructed by us, can be recognized when a dynamic system consists of intrinsic coupled reactions that lead to indefinite evolvability, self-generation, self-maintenance and partial closure and autonomy from the surrounding milieux with which it interacts. |
| Francis Heylighen | A living system is a self-maintaining network of processes that is directed at the goal of preserving its self-maintenance by acting so as to compensate for variations in environmental conditions that would otherwise threaten its self-maintenance. |
| František Baluška | My definition of Life is: Living Cells and All Their Constructs |
| George Ellis | Life is any open system that reproduces and has agency. That is, the way it flexibly interacts with the surrounding environment enhances relative survival prospects of its progeny. |

| Name | Definition |
|---|---|
| Georgi Yordanov Georgiev | Life is a system of any substrate with flows of energy, matter, and information that create efficient internal structure through positive and negative feedback loops and reflects its external environment. It is a part of a network of differentiated similar systems, competing and cooperating, spreading information, developing and evolving exponentially until an external limit is reached. More organized systems are more 'alive'. |
| Gordana Dodig-Crnkovic | Life, from a scientific standpoint, is a natural phenomenon that emerges from the laws of physics and chemistry operating within the universe. It is a process characterized by the self-organization of matter into increasingly complex info-computational networks systems capable of sustaining, reproducing, and adapting themselves to their environment. Life = cognition. |
| Jack Tuszynski | Life is a self-sustained out-of-thermodynamic equilibrium process driven by metabolic energy supply and aimed at self-replication. Importantly, living systems possess various degrees of autonomous movement in response to external conditions but also as an expression of free decision making. |
| James Shapiro | A self-maintaining or reproducing entity that acts purposefully and intelligently to survive but is also capable of self-modification when survival is challenged. |
| Jamie Davies | Living things are non-equilibrium systems, semi-permeable to their environments in ways they mostly control, rich in feedback and with the ability to maintain and repair themselves. Populations of similar living things have the ability to reproduce, though the ability to reproduce is not necessarily a property of every individual living thing. The status of being alive requires organization and action, and this may in principle be realized by a variety of underlying material compositions. |
| Jeremy Gunawardena | I feel that, with our current understanding, defining life is better left to philosophers than to scientists, as it does not seem to lead to experimental or theoretical questions that we can effectively tackle. I prefer to focus on certain properties, such as learning, which can be both formulated theoretically and also lead to experimental approaches that we can currently undertake. Learning does not constitute life, as inanimate entities also learn, but it gets us a lot closer to something resembling life than the current paradigm of selfish genetic machines. |
| Jordi Vallverdú | Life is a self-sustaining process that reduces internal entropy through energy exchange, driven by embodied intentionality to maintain internal order and adapt to its environment, generating value through the interaction between the organism and its world. |
| Joshua Bongard | Information flows are discovered that are initially restricted to narrow spatial and temporal ranges, are univariate (one point source), and the carriers of information are restricted to a few energetic and physical modalities (electricity; vibration). Over time, these flows are found to multiply; expand across space and time; become multivariate and higher-order; and be carried by additional energetic and physical modalities (quantum; plasma). |
| Karl Friston | Life is a scale -invariant, far-from-equilibrium process with an attracting set of characteristic states (technically, pullback attractors) that (i) feature an individuation of living systems in terms of their (Markov) boundaries, (ii) which are open, in the sense there is exchange across the boundaries and (iii) breaks detailed balance, in the sense of possessing solenoidal (i.e., non-dissipative) dynamics; e.g., biorhythms, lifecycles and replication. |
| Kenneth Stanley | If the aim is to define life as it could be, it's a far more subjective endeavor given the vast space for opinion to play a role (e.g. can a computer program be alive, does life require consciousness, etc.). Because of this wobble between the superficial and the subjective, I think it may be more fruitful simply to enumerate candidate configurations that would be particularly interesting for one reason or another, even if no single comprehensive definition is at hand. |

| Name | Definition |
|---|---|
| Kevin Mitchell | A living organism is a self-perpetuating pattern of interlocking dynamical processes, comprising a collective regime of constraints, which does thermodynamic work to persist, far from equilibrium with its surroundings, as a distinct entity through time. Living organisms comprise causal protectorates, and, through the key property of selective historicity, accumulate causal potential and the ability to act as autonomous agents in the world. |
| Larissa Albantakis | In contrast to consciousness (the feeling of being), life is a functional construct which allows for a reductionist definition as a set of sub-processes such as homeostasis, reproduction, etc. Any particular definition proposed is neither right or wrong, but rather more or less specific, inclusive, and above all useful for researchers of various fields. In this view, being "alive" is just a label and therefore ultimately just a matter of definition. |
| Leo Caves | (recursive definition - enter)... Life is a relatively persistent dynamic pattern that is discerned in the coupling of an observer with their environment; the pattern is a manifestation of intrinsically recursive dynamics on a closed topological form in a complementary space of possibility ~ actuality. Resonance of these rhythmic patterns generates (and induces) rich “harmonic” eurhythmically-sustained heterarchical structures, expressed in energy and matter homeorhesis capable of conservative and radical morphodynamics exhibiting a repertoire of autonomous and other rich behaviors; the observer (as such an entity) is complementary to and co-determines its environment—capable of forming the distinction that... (→ re-enter) |
| Mark Solms | A system is alive if it spontaneously and in a thoroughgoing manner strives to oppose the Second Law, to maintain its own existence. |
| Martin Picard | A self-sustaining system capable of healing that uses and organizes energy to achieve expression(s) of consciousness. |
| Michael Levin | We call alive those cognitive systems that (a) self-assemble under metabolic constraints (scarcity of time and energy, which demands essential coarse-graining and mind-full models of self and world, as well as memory reinterpretation that commits to sense-making over fidelity), and (b) are good at scaling up the basal competencies of their parts into a new being with top-down control and a larger cognitive light cone than any of its subsystems. Living beings remember, and anticipate, coarse-graining experience into a continuous actionable dream that forms models of itself and its world, and the distinction between them. Life is a stream of temporarily persistent, coherent, agential patterns within some medium that, over time, expands its cognitive light cone and projects it into new problem spaces within which to strive. |
| Nick Lane | A formal definition of life is unhelpful because living is a process over time, and drawing a line across a continuum is always arbitrary. Geochemical systems that give rise to life combine continuous flow with cellular structure, as in structured hydrothermal systems that operate as electrochemical flow reactors. Environmental disequilibria convert gases such as $H_2$, $CO_2$ and $NH_3$ into organic molecules through a reaction network that prefigures metabolism, from which genes emerge, meaning that growth precedes heredity. |
| Oded Rechavi | Anything that keeps changing in order to stay the same. |
| Olaf Witkowski | Life is hard. A working definition captures it as a physical quine (as coined by Doug Hofstadter): an information pattern capable of self-reproducing through time and space, at various scales, parasitizing the fabric of its substrate. Paradoxically, this makes the pattern both independent of and contingent on this substrate. Life is also soft. We humans can't help but see it as an experiential quine. Sure, the living pattern could be perceived by some third-person party situated "outside the simulation." But we typically encounter observations as third-person observers within the same universe, or by the pattern's own first-person perspective—in the hardware's soft mind's eye. Like the Ship of Theseus, life may, ever so softly, change its own components while maintaining some invariants. Combining the hard and soft pictures above emphasizes the continuity of experience as central to life’s ability to adapt and navigate its universe. |

| Name | Definition |
| --- | --- |
| Paco Calvo | Life is characterized by the possession of intrinsic goals and purposes (phenomenological), a drive for preservation based on interaction-dominant dynamics (e.g., Gaian self-regulation) among biological or artificial components, and the ability to exist within real or virtual environments (not necessarily carbon-based). |
| Pamela Lyon | Life is a process by which a sensitive, behaving system constructs, maintains and reproduces its own organization by taking in and using matter and energy from its surroundings selected on the basis of cognition—awareness, perception, discrimination, valuation, memory, learning—to enable agency, actions aimed at realizing internally generated goals, such as system persistence. |
| Paul Davies | Following Schrodinger, there is a 'new kind of physical law' prevailing in living matter. So we won't have a definition until we uncover that law. What can we say about it in the interim? In my opinion, it would give causal efficacy to information, but above and beyond mere Maxwell demonics. (Life is full of Maxwell demons playing the margins of the second law of thermodynamics.) Following Landauer, 'information is physical,' so it is a quantity that has causal consequences. But living matter couples information networks to chemical networks, and the said 'law' would involve some sort of global measure of complexity. The law would be of a 'new kind' because it would not involve locally-defined quantities. |
| Perry Marshall | Life is the fire that lights itself. Self awareness and self agency - not just an act of creating, but the continual gift of creativity. The potenza for singularity in every single moment, ineffable. |
| Philip Ball | I don't believe we have an adequate understanding of the range of possible behaviors in complex matter to attempt any universal definition of life, but to the extent that we know them, living systems all exhibit agency: the ability to manipulate themselves and their environment to achieve autonomously determined goals. I believe that it should be possible to operationalize a notion of agency more readily than to do so for "life". Characteristic features would include a sustained thermodynamic separation of agent and environment, and a context- and history-dependence of responses to stimuli: an agent's behavior makes reference not just to external conditions but to internal states. In addition: evolution by natural selection is not an intrinsic property of living things, but rather, is the only mechanism currently known by which the necessary complexity for agency might arise spontaneously. |
| Pier Luigi Gentili | 1) Information: The first peculiarity of living beings is that of exploiting matter and energy to gather and utilise Information to pursue goals. Every living being has a hierarchical structure, and different types of information are encoded at various levels<br>2) Life Cycle: Every living being is an open system characterised by a boundary that delimits it from the rest of the environment. It has a birth, a life during which it grows and ages. It can self-maintain, self-reproduce, and protect itself from intruders and harmful elements. However, finally, it dies when all fundamental internal activities cease.<br>3) Adaptation-Acclimation-Evolution: During its lifetime, every living being can adapt by adjusting its metabolic processes, acclimating by expressing new genes, and evolving by changing its genome under an ever-changing environment. |
| Pranab Das | LIFE is the collection of all systems capable of ongoing self-assembly, regulation and endurance that exceeds by many orders of magnitude the characteristic time periods of their internal function. Living things are instantiations of a continuing set of processes which can be understood as a "story" carried down and reproduced through the generations. |
| Predrag Slijepčević | My starting assumption is that life – the totality of the living world – is always more complex than our models of life. My second assumption is that life is a non-ergodic evolving system – a system in a permanent state of becoming. My third assumption, following from the second one, is that life is a processual system that cannot be fully described by standard physicalist approach because it is blind to a deep (evolutionary) future and it ignores extra-terrestrial forms not based on standard biochemistry or biophysics. |

| Name | Definition |
|---|---|
| Ricard Solé | Life is a dynamic, self-sustaining collective property emerging by exploiting self-organization of soft matter in cooperation with information-carrying substrates. These features include encapsulation as a precondition to achieving agency and might not be associated with replication. |
| Richard Watson | Life is the pattern and process of love – a deeply vulnerable mutual dance. It is created by mutually transformative connection (not separation, or survival or self-interest), and the greater the depth of vulnerability, letting go of ego that blocks genuine connection, the more life is lived. The song of life that orchestrates the dance arises from the sustained harmonic vibration between parts and wholes (both within self and in larger context), that carries the resonance of its entire history of experience, and in so doing, repeatedly recreates the conditions for its own origination and animation, materialising the singer that continues the living transformation of the song. |
| Rita Pizzi | Complex agglomerate of carbon-based elements, with self-organizing and self-replicating properties. The base might not be carbon in other parts of the universe. Self-organizing is not enough; what distinguishes life from other self-organizing systems is the presence of DNA, i.e. an organic macromolecule capable of containing all the information to reproduce the life form that contains it. |
| Scott Gilbert | A living entity is a bounded system of molecular processes that interact with its environment so as to use metabolism to persist. The living entity would therefore retain its identity by changing its parts |
| Sonia Sultan | Life describes the ability of a naturally occurring system to continuously perceive relevant aspects of its external environment, and to build and maintain itself through continuous yet finite, real-time processes of response to those perceived conditions. In my view the key components are environmental perception, response, and the combination of those two continuous activities with the system's finitude. |
| Steve Frank | In general, I believe it is almost always a bad idea to try and define something outside the context of solving a specific puzzle or working toward a specific goal. That is like deciding which tools to use before you know what you are trying to build or fix because definitions are tools and not endpoints. That is the reason that people never agree on definitions and argue endlessly and uselessly about them, there are no good definitions outside of context. |
| Stuart Kauffman | We have no agreed-upon definition of life. We here build toward the following: Life is a non-equilibrium, self-reproducing chemical reaction system that achieves: i) Collective autocatalysis, ii) Constraint Closure, iii) Spatial Closure, iv) As such, living entities are Kantian Wholes. |
| Stuart Newman | Life is a form of organized matter out of thermodynamic equilibrium, capable of maintaining its organized state by selective material exchange with its external environment. It is entirely continuous with the nonliving world at the level of atomic elements, but many molecules essential to their maintenance are uniquely produced by living systems. The principles underlying the characteristic organization of living systems have so far not been identified in the nonliving natural world. |
| Sui Huang | Material entities in the universe, such as atoms, molecules, cells, organisms, which are variants of larger sets with shared common properties are prone to interact in some rule-constraint manner, following elementary physical laws. In doing so, they create through constrained combinatorial explosion an immense diversity of larger entities, governed by more rules that appear increasingly detached from the elementary physical interactions: biology. Subjugated to these rule-like constraints, these entities are prevented from direct descend along the free energy gradient imposed by thermodynamics but instead, are stuck far away from thermodynamic equilibrium, akin to water hold up by a dam, and, consuming free energy, they follow instead the laws of dynamical systems and produce behaviors of particular patterns, among which are cycles of replication that we mistake for something non-physical: life. |

| Name | Definition |
| --- | --- |
| Susan Stepney | Living systems are open far-from-equilibrium self-producing systems, embodied in smart computational metamaterials, with non-trivial, irreversible meta-dynamics (where metamaterials are engineered or evolved to have special functional and/or computational behaviors, and meta-dynamics is the dynamics, or change, of the system's own dynamics). |
| Theodore Pavlic | Living things are statistical novelty producers, and sometimes the products of life (viruses, computers, robots) can themselves be extensions of life that are understandably confused for the genuine article. Novelty production is really what all of the other putative necessary features of life (metabolism, growth, organization, homeostasis, reproduction, etc.) bring about. On another planet, it won't be any particular chemical process or molecular structure that will really convince us that we have found life; it will be the repeated presence of patterns that simple historical contingency has no chance of producing continuously at scale. |
| Timothy Jackson | Life as we know it is a particular evolutionary lineage of self-organising systems. Self-organising evolution appears to be generic, but the life lineage has achieved its peculiar degree of diversity and complexity due to its unique mechanisms of information storage and retrieval. These mechanisms function as constraints on ubiquitous processes and either require (e.g., metabolism) or enable (e.g., agency) some of those properties commonly associated with life. Technical objects (tools, machines, etc) constructed by living systems are best conceived of as internal to the life lineage, as components of the extended phenotypes of the organisms that produce them. |
| Tom Froese | Biological organization makes use of physical tendencies toward efficient energy flows by diverting them into a process of individuation. But this systemic imposition of constraints is not sufficient for flexible behavior, which requires the injection of a factor of arbitrariness into state dynamics. These three complementary roles, (1) energy flow maintenance, (2) systemic identity creation, and (3) state constraint destruction, suggest a complex ontology of the organism characterized by three distinctive domains: matter, life, and mind, respectively. |
| Walter Fontana | Life is matter that has learned how to learn. |
| Wesley Wong | An operational definition of life, grounded in physical measurement, could help us recognize life in unexpected contexts. While no single definition covers all known phenomenon, approaches like Schrodinger's definition based on statistical mechanics, and Laughlin's framing of emergent behavior in mesoscopic systems are particularly compelling. In the end, we may not need a singular definition--instead, context-specific definitions, including those that blur the line between life and non-life, could be more useful. |
| Will Ratcliff | A self-sustaining chemical system capable of open-ended Darwinian evolution. |
| William Miller | Cognitive decision-making and problem-solving as self-referential information processing and management. |

***Supplemental Table 2:*** *Correlation Matrix Statistics*

| Model | Correlation Range | Mean | Median | Standard Deviation | Skewness |
|---|---|---|---|---|---|
| Claude 3.7 Sonnet | -0.80 to 0.83 | 0.30 | 0.37 | 0.372 | -0.972 |
| Llama 3.3 70B Instruct | -0.90 to 0.85 | 0.17 | 0.20 | 0.510 | -0.903 |
| GPT-4o | -0.85 to 0.97 | 0.20 | 0.32 | 0.397 | -0.756 |
| Multi-model Integration | -0.84 to 0.82 | 0.22 | 0.32 | 0.394 | -0.740 |

***Supplemental Table 3:*** *Cross-model Clustering Statistics*

| Model | Number of Clusters | Largest Cluster Size | Intra-cluster Correlation | Inter-cluster Correlation | Contrast Metric |
|---|---|---|---|---|---|
| Claude 3.7 Sonnet | 7 | 25 | 0.495 ± 0.225 | 0.237 ± 0.389 | 0.258 |
| Llama 3.3 70B Instruct | 11 | 32 | 0.516 ± 0.240 | -0.047 ± 0.548 | 0.563 |
| GPT-4o | 11 | 16 | 0.548 ± 0.172 | 0.140 ± 0.394 | 0.408 |
| Multi-Model Integration | 8 | 24 | 0.509 ± 0.187 | 0.150 ± 0.390 | 0.359 |

***NOTE*** All supplemental images are also available in the Github source code repository here: [https://github.com/mrbende/what-lives/tree/main/supplemental](https://github.com/mrbende/what-lives/tree/main/supplemental)

***Supplemental Figure 1:*** *Definitions of Life Correlation Matrix;* ***Claude 3.7 Sonnet***

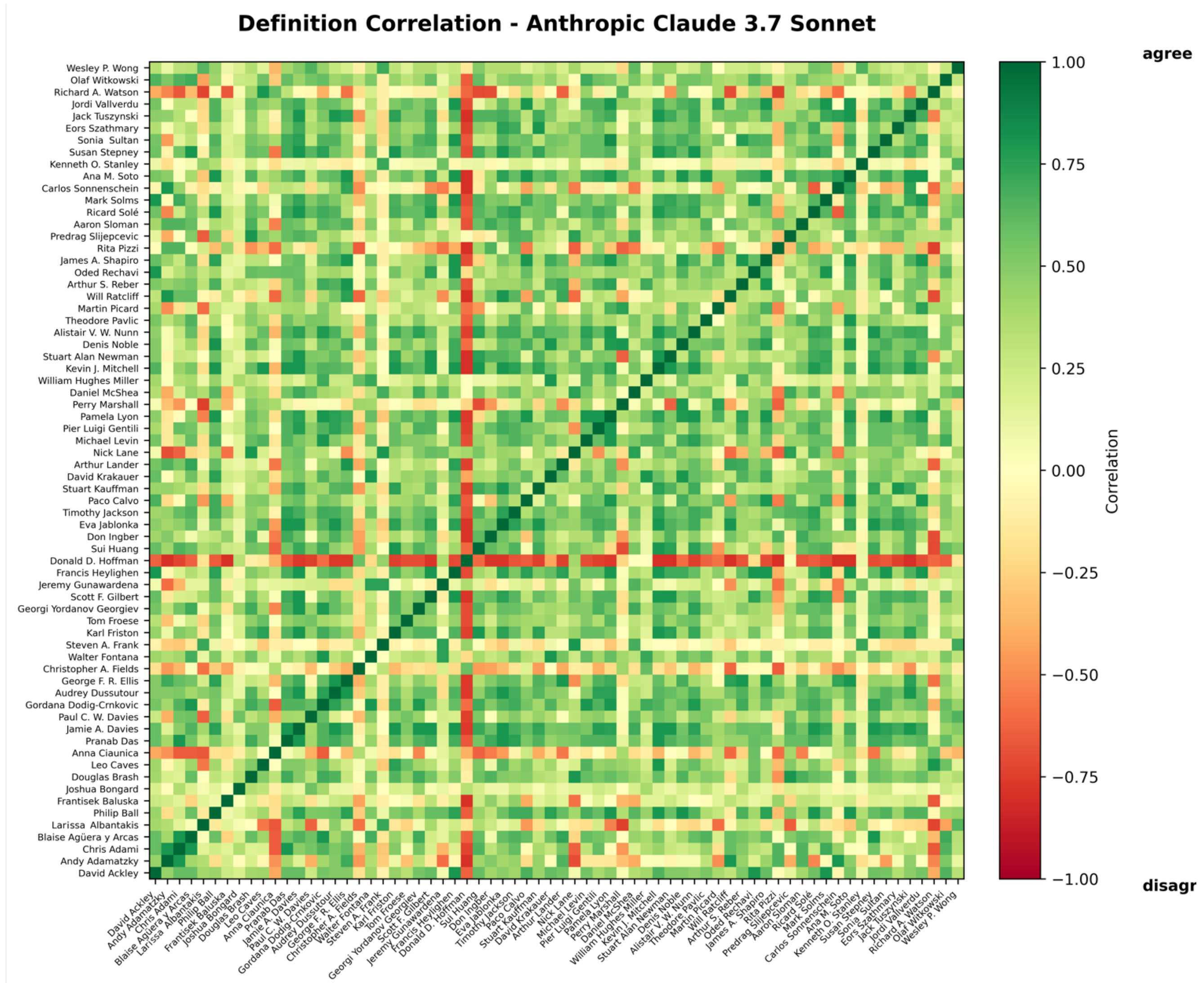

***Supplemental Figure 2:*** *Definitions of Life Correlation Matrix;* ***Llama 3.3 70B Instruct***

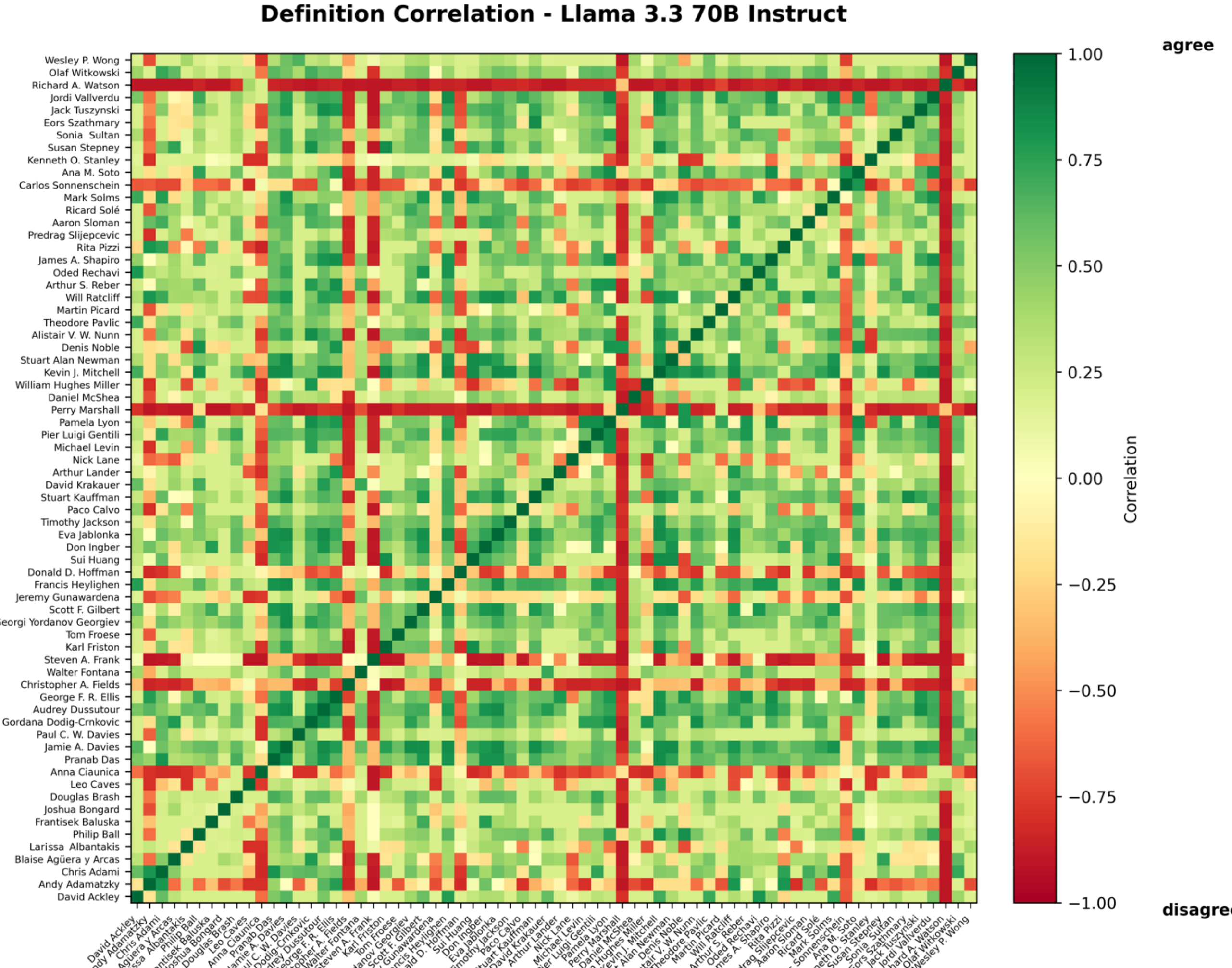

***Supplemental Figure 3:*** *Definitions of Life Correlation Matrix;* ***GPT-4o***

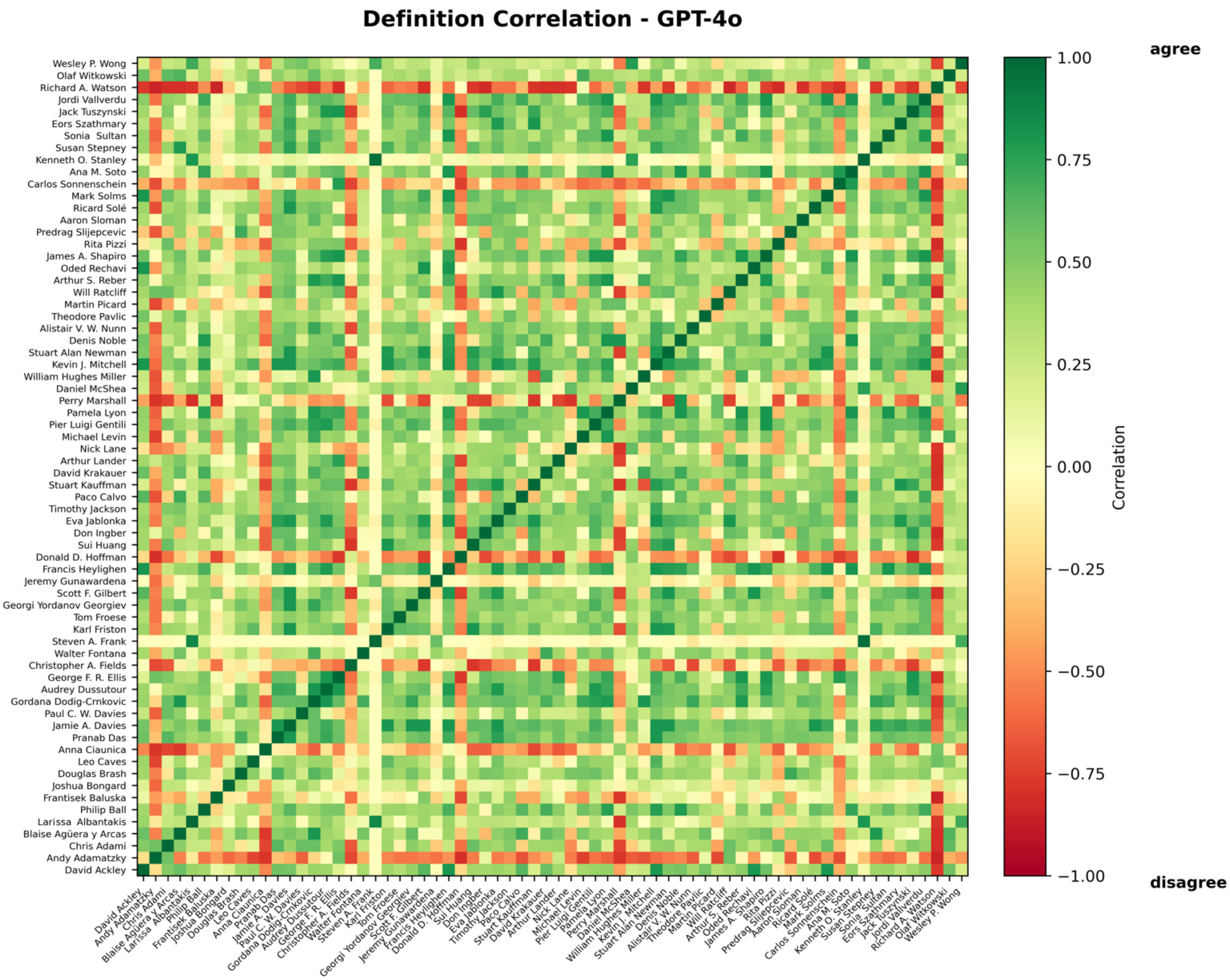

***Supplemental Figure 4:*** *Definitions of Life Correlation Matrix; ;* ***Averaged Claude 3.7 Sonnet / Llama 3.3 70B Instruct / GPT-4o****; symmetrized average*

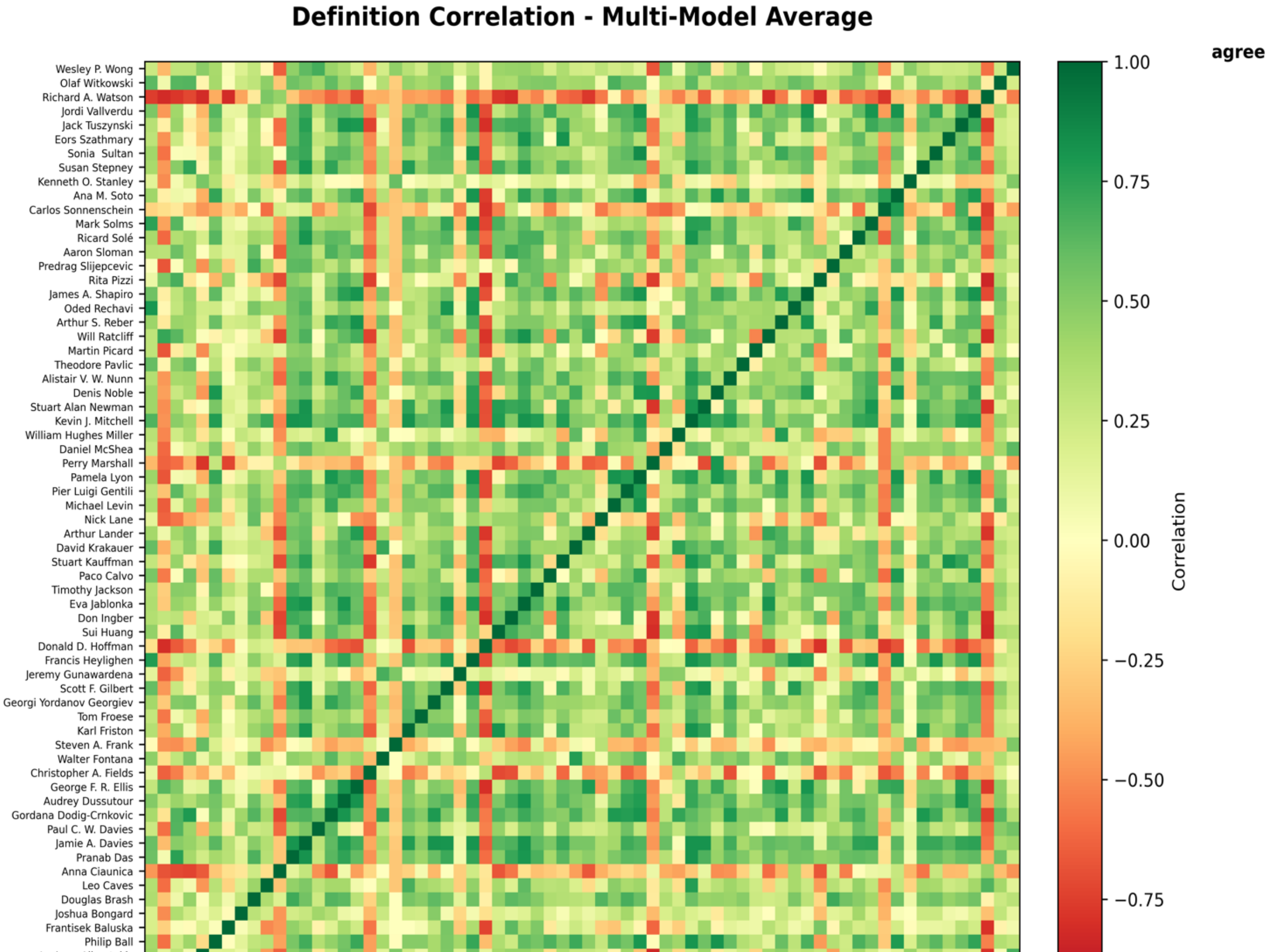

***Supplemental Figure 5:*** *Definitions of Life Entropy Matrix;* ***Claude 3.7 Sonnet***

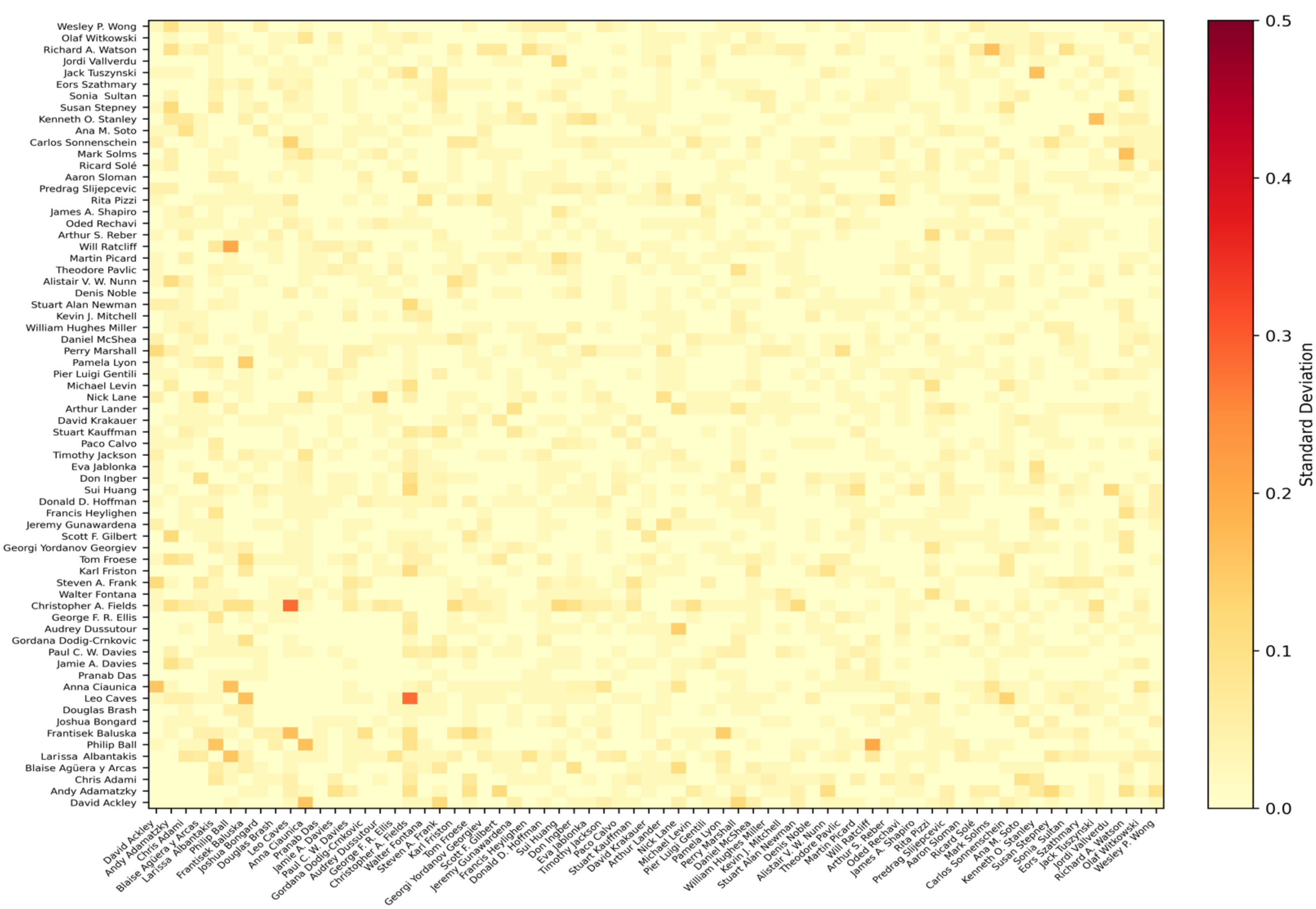

***<u>Supplemental Figure 6:</u>*** *Definitions of Life Entropy Matrix;* ***Llama 3.3 70B Instruct***

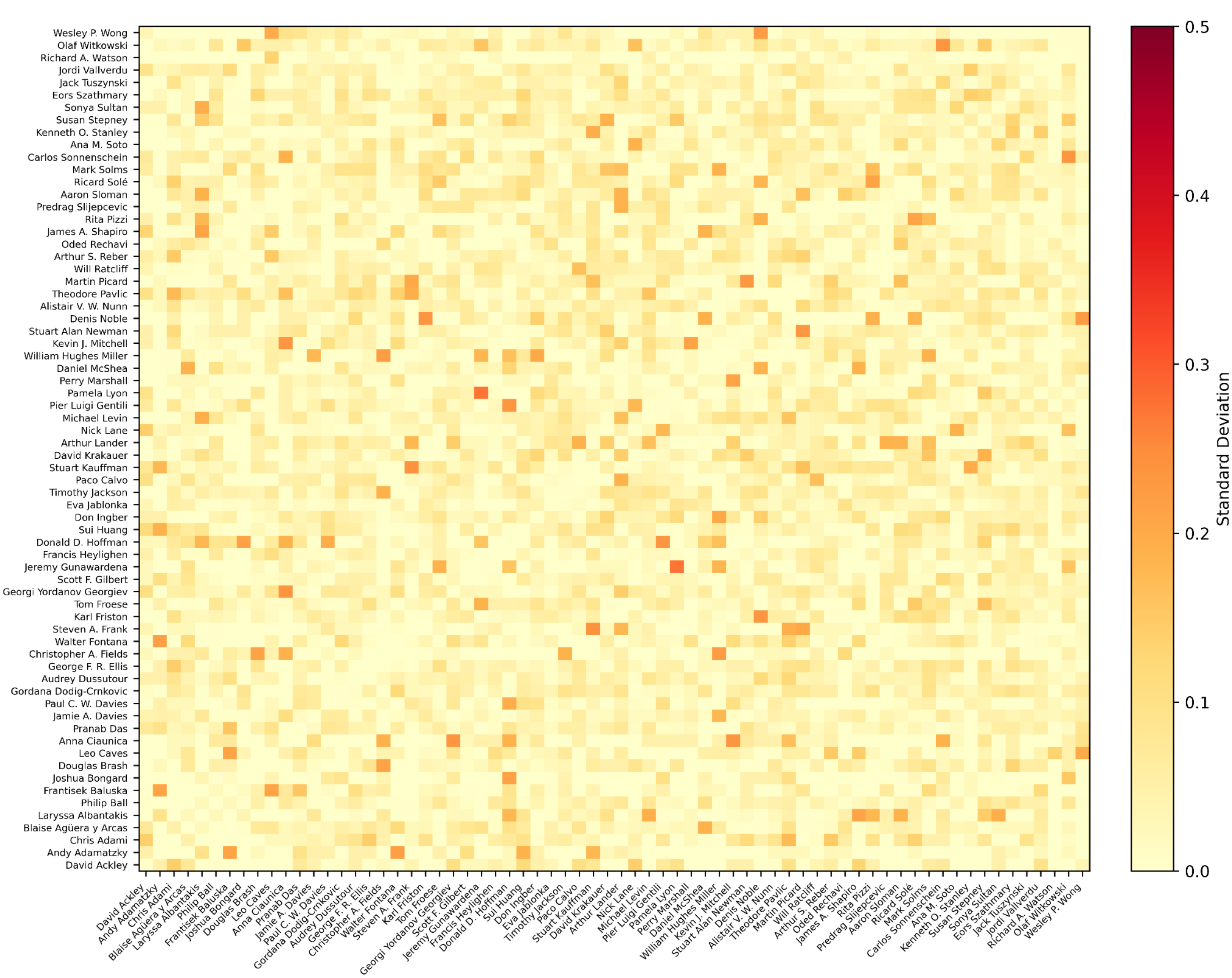

***<u>Supplemental Figure 7:</u>*** *Definitions of Life Entropy Matrix;* ***GPT-4o***

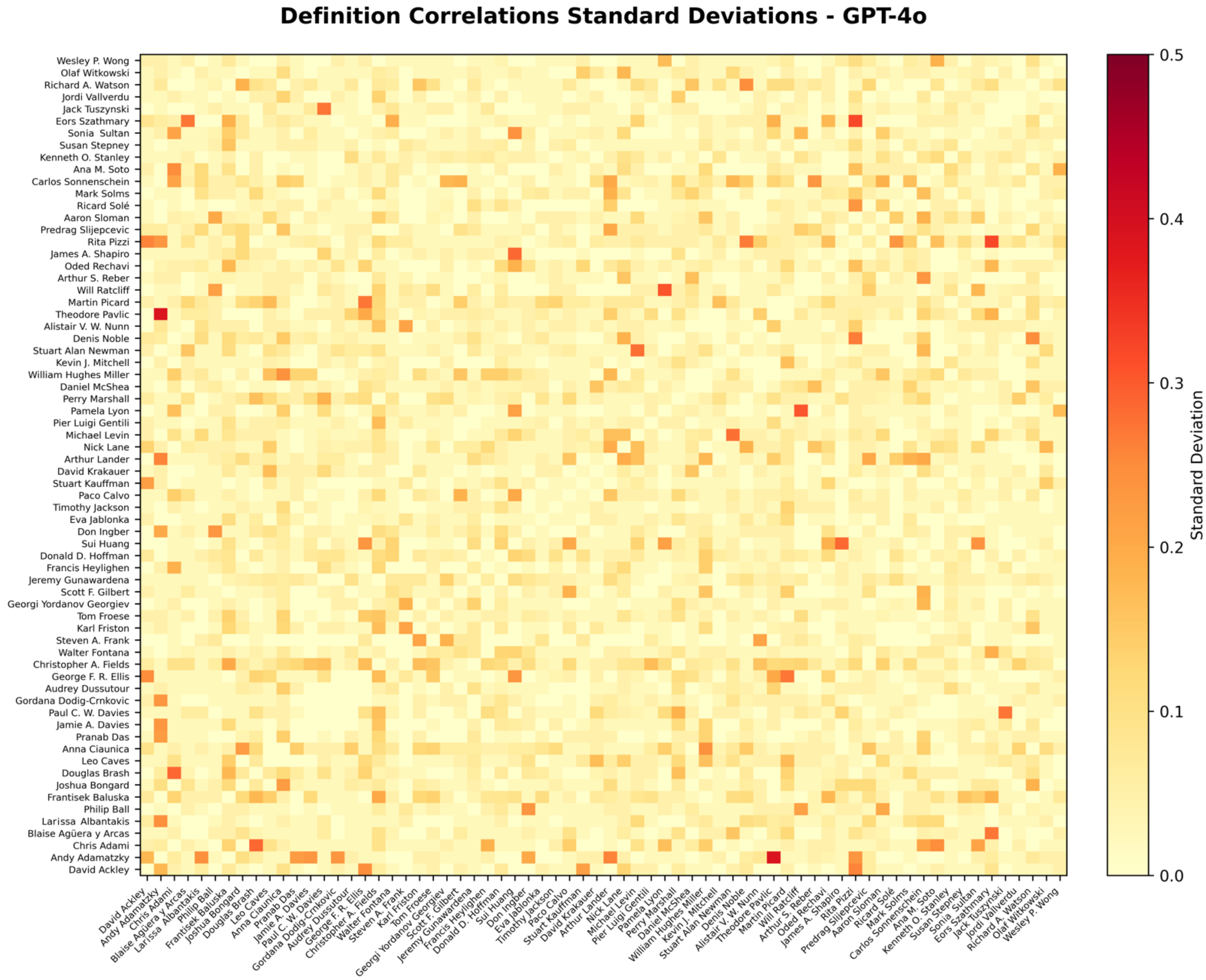

***Supplemental Figure 8:*** *Clustered and Sorted Correlation Matrices of Life Definitions;* ***Claude 3.7 Sonnet***

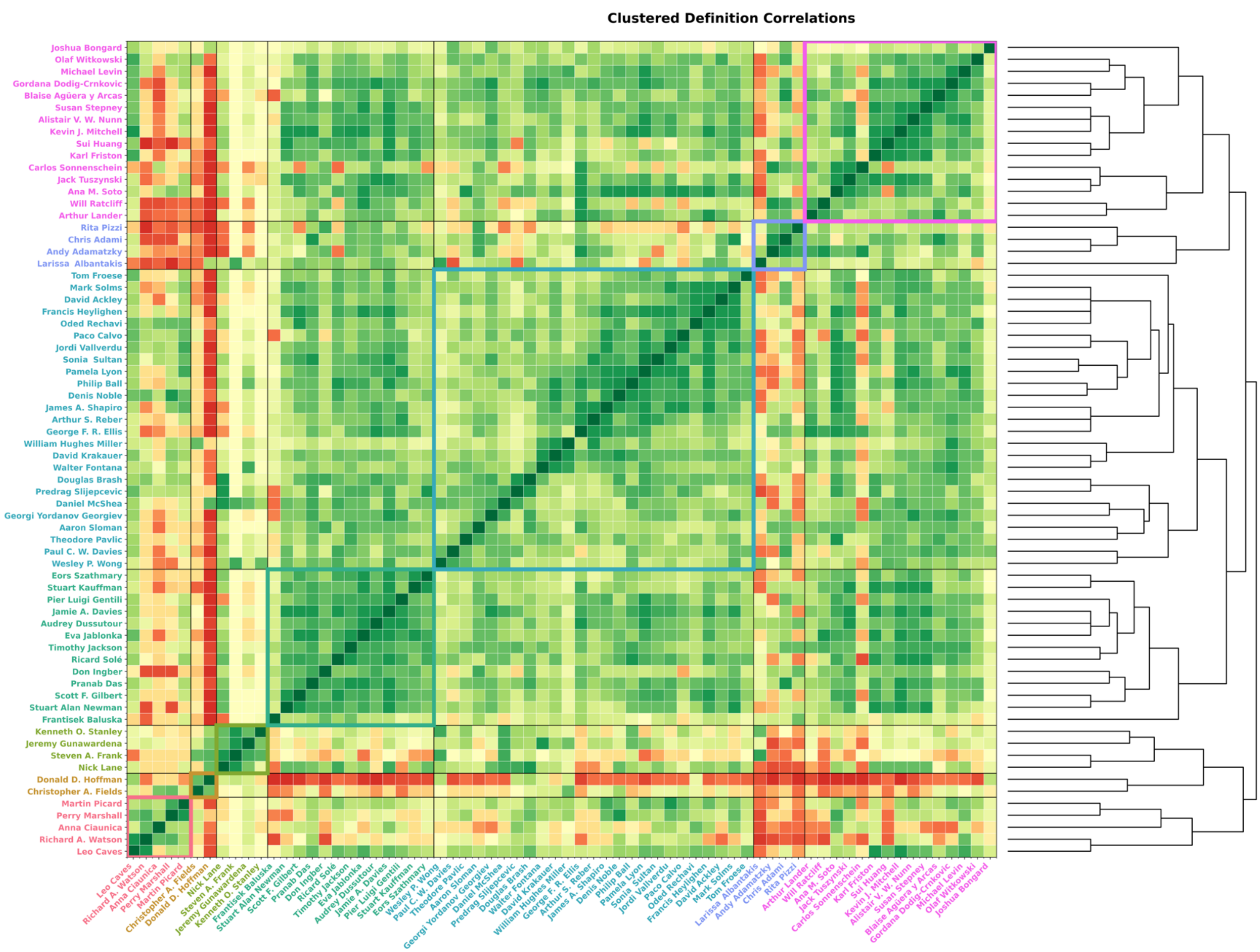

***Supplemental Figure 9:*** *Clustered and Sorted Correlation Matrices of Life Definitions;* ***Llama 3.3 70B Instruct***

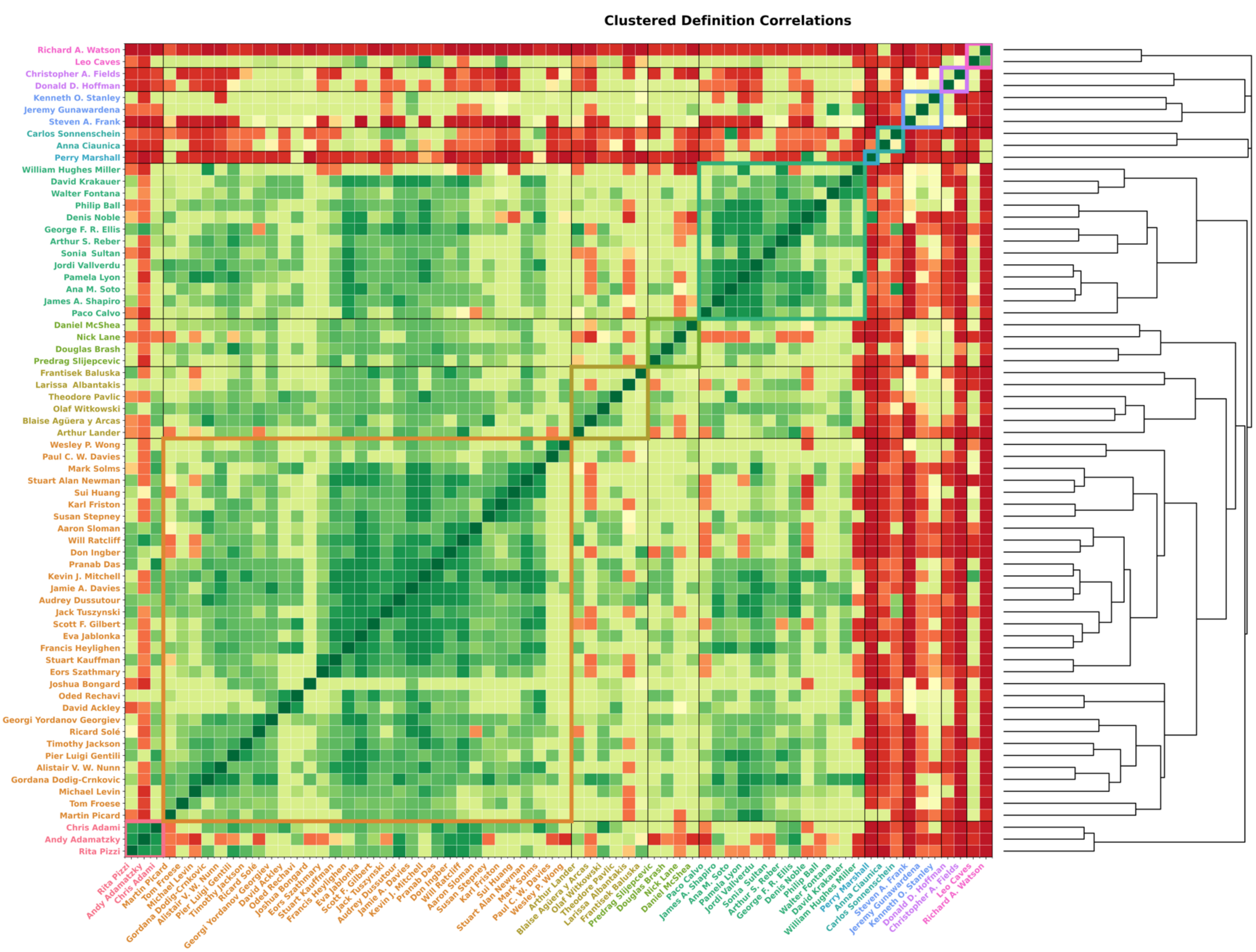

***Supplemental Figure 10:*** *Clustered and Sorted Correlation Matrices of Life Definitions;* ***GPT-4o***

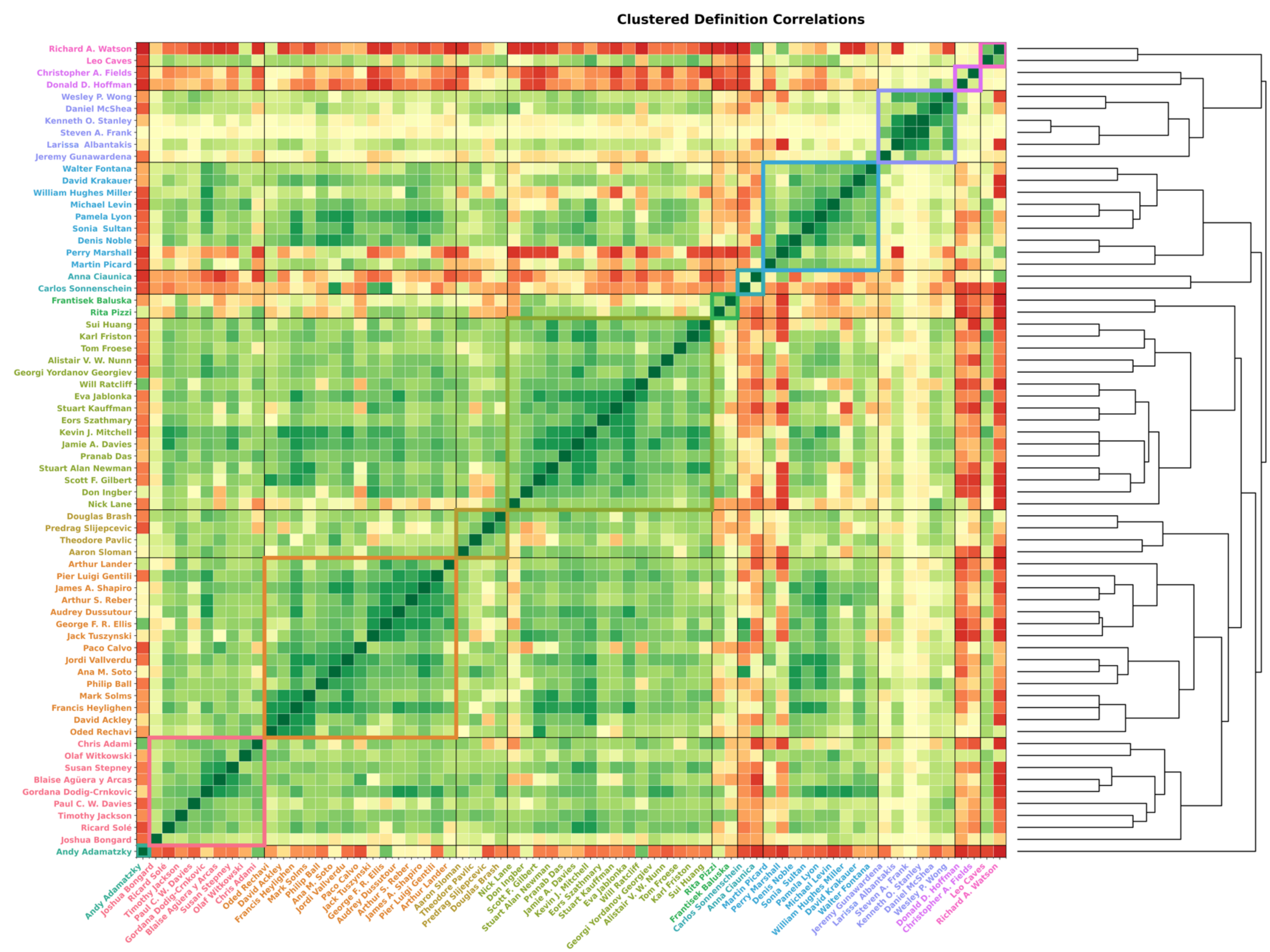

***Supplemental Figure 11:*** *2D Feature Projection of Life Definitions – Panel of Clustering Methods*

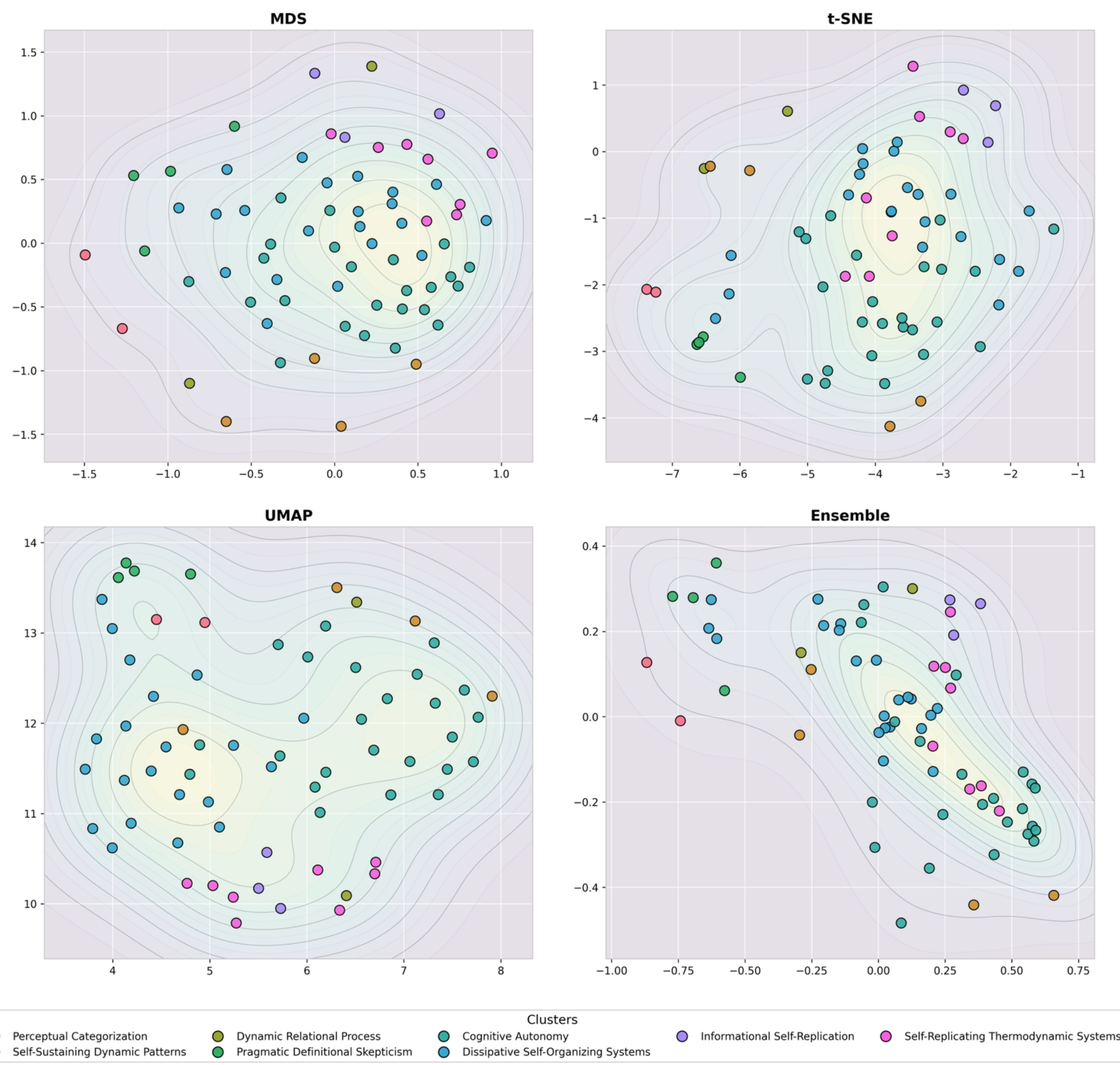